\documentclass[12pt]{article}

\usepackage{a4}
\usepackage{epsfig}

\newif\iffigframe
\newif\ifbigfig
\bigfigtrue         % bigger version of figures at end of paper

\makeatletter
\newcommand{\mkfig}[5]{%
  \ifbigfig
    \expandafter\def\expandafter\mk@figs\expandafter{\mk@figs
      \begin{figure}[p]%
        \hbox to\hsize{\hss\figbox{#3\hsize}{#4}{#1}\hss}%
        \caption[#1]{\label{#1}#5}%
      \end{figure}}%
  \else
    \begin{figure}[ht]%
      \hbox to\hsize{\hss\figbox{#2\hsize}{#4}{#1}\hss}%
      \caption[#1]{\label{#1}#5}%
    \end{figure}%
  \fi}
\def\mk@figs{}
\def\mkfigs{\mk@figs\def\mk@figs{}}
\makeatother

\newcommand{\figbox}[3]{\hbox to#1\bgroup
  \dimen0=1bp \dimen1=#1\relax
  \def\a##1 ##2 ##3 ##4 ##5\\{\if!##1!\a##2 ##3 ##4 ##5 .\\\else
    \dimen3=##3\dimen0 \advance\dimen3 -##1\dimen0
    \dimen4=##4\dimen0 \advance\dimen4 -##2\dimen0
    \dimen5=\dimen4 \divide\dimen5 \dimen3
    \dimen2=\dimen1 \multiply\dimen2 \dimen5
    \multiply\dimen5 \dimen3 \advance\dimen4 -\dimen5
    \dimen5=\dimen1
    \loop \advance\dimen4 \dimen4 \divide\dimen5 2
    \ifnum\dimen5>0 \ifnum\dimen4<\dimen3 \else
      \advance\dimen4 -\dimen3 \advance\dimen2 \dimen5 \fi
    \repeat
    \dimen5=10\dimen1 \divide\dimen5 \dimen0
    \includegraphics{#3.eps}%
    \iffigframe \vrule\hss \else \hfil \fi
    \vbox to\dimen2\bgroup
      \iffigframe \hrule width\dimen1\vss \hrule \else \vfil \fi
      \egroup
    \iffigframe \vrule\hss \fi
    \egroup\fi}%
  \a#2 . . . .\\}

\newcounter{subequation}[equation]

\makeatletter
\expandafter\let\expandafter\reset@font\csname reset@font\endcsname
\newenvironment{subeqnarray}
  {\arraycolsep=1pt
    \def\@eqnnum\stepcounter##1{\stepcounter{subequation}{\reset@font\rm
      (\theequation\alph{subequation})}}\eqnarray}%
  {\endeqnarray\stepcounter{equation}}
\makeatother

\newcounter{statement}
\newenvironment{statement}[4]
  {\par\vspace{1mm}\refstepcounter{statement}
    \noindent#1#2 \arabic{statement} #4\unskip: #3}{\par\vspace{1mm}}

\newenvironment{statement*}[4]
  {\par\noindent#1#2 #4\unskip: #3}{\par\vspace{2mm}}

\newcommand{\PrfMark}{\leavevmode\vrule width1.5exheight1.5exdepth0pt}

\begin{document}

\date{}
\newcommand{\dd}{\mbox{d}}
\newcommand{\tr}{\mbox{tr}}
\newcommand{\la}{\lambda}
\newcommand{\ka}{\kappa}
\newcommand{\al}{\alpha}
\newcommand{\ga}{\gamma}
\newcommand{\de}{\delta}
\newcommand{\si}{\sigma}
\newcommand{\bomega}{\mbox{\boldmath $\omega$}}
\newcommand{\bsi}{\mbox{\boldmath $\sigma$}}
\newcommand{\bchi}{\mbox{\boldmath $\chi$}}
\newcommand{\bal}{\mbox{\boldmath $\alpha$}}
\newcommand{\bpsi}{\mbox{\boldmath $\psi$}}
\newcommand{\brho}{\mbox{\boldmath $\varrho$}}
\newcommand{\beps}{\mbox{\boldmath $\varepsilon$}}
\newcommand{\bxi}{\mbox{\boldmath $\xi$}}
\newcommand{\bbeta}{\mbox{\boldmath $\beta$}}
\newcommand{\ee}{\end{equation}}
\newcommand{\eea}{\end{subeqnarray}}
\newcommand{\be}{\begin{equation}}
\newcommand{\bea}{\begin{subeqnarray}}
\newcommand{\ii}{\mbox{i}}
\newcommand{\e}{\mbox{e}}
\newcommand{\pa}{\partial}
\newcommand{\Om}{\Omega}
\newcommand{\vep}{\varepsilon}
\newcommand{\bfph}{{\bf \phi}}
\newcommand{\lm}{\lambda}
\def\theequation{\arabic{equation}}
\renewcommand{\thefootnote}{\fnsymbol{footnote}}
\newcommand{\R}{{\rm I \hspace{-0.52ex} R}}
\newcommand{\N}{{\sf N\hspace*{-1.0ex}\rule{0.15ex}%
{1.3ex}\hspace*{1.0ex}}}
\newcommand{\Q}{{\sf Q\hspace*{-1.1ex}\rule{0.15ex}%
{1.5ex}\hspace*{1.1ex}}}
\newcommand{\C}{{\sf C\hspace*{-0.9ex}\rule{0.15ex}%
{1.3ex}\hspace*{0.9ex}}}
\newcommand{\eins}{1\hspace{-0.56ex}{\rm I}}
\renewcommand{\thefootnote}{\arabic{footnote}}

\title{\hbox to0.9\hsize{\strut}
Regular solutions to higher order curvature
Einstein--Yang-Mills systems in higher dimensions}
\author{{\large Peter Breitenlohner,}$^{\ddagger}$
{\large Dieter Maison}$^{\ddagger}$
and {\large D. H. Tchrakian}$^{\dagger \star}$ \\ \\
$^{\ddagger}${\small Max-Planck-Institut f\"ur Physik}\\
{\small Werner-Heisenberg-Institut}\\
{\small F\"ohringer Ring 6, D-80805 M\"unchen, Germany}\\ \\
$^{\dagger}${\small Department of
Mathematical Physics, National University of Ireland Maynooth,} \\
{\small Maynooth, Ireland} \\
$^{\star}${\small School of Theoretical Physics -- DIAS, 10 Burlington
Road, Dublin 4, Ireland }}

\date{}

\maketitle

%\ \ \ PACS Numbers: 04.50.+h, 11.10.Kk, 11.15.Kc

\bigskip

\begin{abstract}
We study regular, static, spherically symmetric solutions of Yang-Mills
theories employing higher order invariants of the field strength coupled to
gravity in $d$ dimensions. We consider
models with only two such invariants characterised by integers $p$ and
$q$. These models depend on one dimensionless parameter $\alpha$ leading to
one-parameter families of regular solutions, obtainable by numerical
solution of the corresponding boundary value problem. Much emphasis is put
on an analytical understanding of the numerical results.

\end{abstract}

%%%%%%%%%%%%%%%%%%% put preprint number at topright %%%%%%%%%%%%%%%%%%%
\vfill
\vbox to 0pt{
    \vskip-\vsize
    \hbox to\hsize{\hfil\vbox{%
        \hbox{MPP-2005-83}%
        \hbox{August 8, 2005}%
    }}
    \vfill
}
%%%%%%%%%%%%%%%%%%%%%%%%%%%%%%%%%%%%%%%%%%%%%%%%%%%%%%%%%%%%%%%%%%%%%%%
\newpage

\section{Introduction}\label{secti}
Regular, static, and finite energy solutions to the Einstein--Yang-Mills
(EYM) equations have presented a subject of interest since the work of
Bartnik and McKinnon~\cite{BM} (BM) in $4$ spacetime dimensions. Besides the
regular solutions also corresponding black hole solutions were
found ~\cite{VG,B,Kuenzle} representing novel examples of
{\it hairy black holes}. In addition, these classical solutions are of
interest also as the `solitons'
of the low energy effective actions of superstring theory~\cite{GSW},
since both gravitational and non Abelian gauge fields occur in the latter and
in gauged supergravities. More specifically, such solutions (especially the
black holes) have an important role to play in quantum gravity~\cite{HS}, and
also on (higher dimensional) $D$-branes~\cite{Pol}. In this wider context
therefore, the construction of such solutions in spacetimes of dimensionality
higher than $d=4$ is of actual interest. This is the purpose of the present
work, namely the study of EYM systems in arbitrary spacetime dimensions $d$.

Although the YM theory in flat space has
no `soliton' type solutions due to its scaling properties, the
coupling to gravity changes the picture. Bartnik and McKinnon \cite{BM}
discovered a discrete family of static, spherically symmetric globally
regular solutions. However, the scaling properties of the EYM theory
expressed by a `Virial Theorem' forbid such solutions for $d\ge5$;
while the scaling dimension of static gravitational fields is $d-3$ the one
of the YM fields is equal to $d-5$. The virial theorem implies that
non-trivial regular solutions require terms in the action with scaling
dimensions of different sign.
The way out proposed in \cite{BCT,BCHT} is to use higher order invariants
for gravity and\slash or YM fields.
The choice of models boils down, in principle, to the selection of various
terms in the gravitational and non Abelian hierarchies of increasing orders in
the respective curvatures, namely the Riemann and the YM curvatures, which are
reparametrisation and gauge invariant. There is an important restriction,
namely that one considers only those Lagrange densities that are
constructed from antisymmetrised $2p$ curvature forms, and exclude all other
powers of both Riemann and YM curvature $2$-forms. (In the gravitational
case this results in the familiar Gauss--Bonnet Lagrangean.) As a result, only
{\it velocity--squared} fields appear in the Lagrangean, which is what is
needed for physical reasons.
Here we will add only the minimal number of such higher order terms that
are necessitated by the requirements of the virial theorem.
This criterion makes the inclusion of higher order gravitational terms
unnecessary and  in fact the inclusion of higher order gravitational terms does
not seem to alter the qualitative properties of the classical solutions \cite{BCT}.
From the viewpoint of classical solutions, indeed, such terms seem redundant.
They become, however, important in the
mathematically interesting situations where the first member of the
YM hierarchy, namely the usual YM term, is absent.
The situation concerning higher order YM curvature terms in the string
theory effective action is complex and as yet not fully resolved. While
YM terms up to $F^4$ arise from (the non Abelian version of) the
Born--Infeld action~\cite{Tseytlin}, it appears that this approach
does not yield all the $F^6$ terms~\cite{BRS}.
Terms of order $F^6$ and higher can also be obtained by employing the
constraints of (maximal) supersymmetry~\cite{CNT}. The results
of the various approaches are not identical.

Thus we end up adding only higher order YM terms to the usual EYM Lagrangean.
The YM hierarchy employed is
\begin{equation}
\label{YMhier}
{\cal L}_P=\sum_{p=1}^{P}\frac{\tau_p}{2(2p)!}\ \mbox{Tr}\,F(2p)^2\;,
\end{equation}
$F(2p)$ denoting the $p$ fold totally antisymmetrised products
\begin{equation}
F(2p)\equiv F_{\mu_1\mu_2...\mu_{2p}}=F\wedge F\wedge\ldots\wedge F\;,
\qquad
p\ {\rm times}\;,
\end{equation}
of the YM
curvature, $F(2)=F_{\mu\nu}$, in this notation. Clearly, the highest value $P$
of $p$ in Eq.~(\ref{YMhier}) is finite and depends on the dimensionality $d$ of
the spacetime. To complete the definition of the models (\ref{YMhier}) the
gauge group $G$ must be specified.
With our aim in the present paper, of constructing
static spherically symmetric solutions in $d$ spacetime dimensions, the
smallest possible such gauge group is $G=SO(d-1)$ requiring $d>4$.

In \cite{BCT,BCHT} the simplest possibility $P=2$ was chosen.
The scaling dimension of the $p=2$ term is equal to $d-9$, yielding a
negative contribution in the virial theorem for $d<9$.
As a consequence these gravitating YM models in spacetime dimensions
$d=5$, $6$, $7$, and $8$ were found to
support particle like solutions%
\footnote{Taking only the $p=2$
term one finds generalisations of the BM solutions in these
dimensions}.
Here we will consider more generally models
involving two arbitrary terms of the sum in Eq.~(\ref{YMhier}), say
$F(2p)$ and $F(2q)$. For simplicity let us call them $p$-$q$-models.
Like the 1-2-model these simplest nontrivial models possess one
dimensionless parameter $\alpha$ providing a measure of the strength of the
gravitational self-interaction of the YM configurations.

The regular solutions reported in \cite{BCT,BCHT}, which were constructed
numerically, featured numerous novel and exotic properties, some of which
were not understood or explained at a more fundamental level. The purpose
of the present work is specifically to understand these
properties analytically, and then substantiate the conclusions and
predictions
numerically. In fact, our analysis immediately generalises to the general
$p$-$q$-models, all showing the same basic structure. In this context it also
turns out to be useful to interpolate in the dimension, i.e.\ take the
dimension parameter $d$ to be an arbitrary real number. This is possible
since the reduced one dimensional equations of motion for static,
spherically symmetric configurations depend parametrically on $d$.

The properties of the solutions alluded to in \cite{BCT,BCHT} are:
\begin{enumerate}
\item
The solutions form one parameter families parametrised by
$\alpha$. It was found that $\alpha$ increases from the flat limit zero up to
some $\alpha_{\rm max}$ and then turns back.
\item
In $d=6$, $7$, and $8$, where there is a finite-energy gravity-decoupling solution,
{\bf two} solutions appear to exist for most values of $\alpha$, up to
$\alpha_{\rm max}$. Starting from the flat $F(2)+F(4)$ solution $\alpha$
increases from zero to a maximal value $\alpha_{\rm max}$ and then runs back to
zero for $d=8$, while for $d=6$ and~$7$ the families end at some finite
non-zero value of $\alpha$
\item
In $d=5$ there is no flat $F(2)+F(4)$ solution, but
the relevant flat solution is the instanton of the pure $F(2)$
theory considered as a static solution of the $d=5$ theory.
After a suitable rescaling both gravity and the $F(4)$ term decouple in the
limit $\alpha\to 0$.
Again $\alpha$ turns back after reaching some maximal value, but then seems
to converge to some limiting critical value $\alpha_{c}$ after performing
a finite or infinite number of oscillations.
\item
Unlike in the case of $d=4$ EYM theory no multi-node solutions were found.
\end{enumerate}

In the present paper we will try to clarify the findings desribed above
with particular emphasis on the critical values of the parameter
$\alpha$ marking the endpoint of the one-parameter families of solutions.
Based on experience with other systems like self-gravitating magnetic
monopoles \cite{BFM1} we propose that the critical solutions run into
certain fixed points (f.p.) of the dynamical system represented by the reduced
field equations, a system of nonlinear ordinary differential equations.
Employing the standard method of linearisation at the f.p.\ it is possible
to study the analytical properties of solutions coming close to the f.p.\
and derive asymptotic expressions for their parameters. Constructing
solutions starting at the f.p.\ itself provides a direct method to
determine critical parameters like $\alpha_{c}$ yielding much more precise
values than extrapolation from solutions coming only close to it.
In fact, our analysis leads to a rather clear picture of the results
listed above.

Concerning point $2.$, ($d=6$, $7$, and~$8$) we claim that the upper branch of the
solution family runs back all the way to $\alpha=0$, the limit
(after suitable rescaling decoupling the $F(2)$ part)
being the analogue of the BM solution for the $F(4)$ model.
As $\alpha$ tends back to zero the solutions for $d=6$ and $d=7$
come very close to a fixed point already known from
the EYM theory in $d=4$ related to the behaviour at the horizon of the extremal
Reissner-Nordstr{\o}m black hole \cite{BFM,BFM1}. Our analytical analysis,
however, excludes a bifurcation with solutions running into the
fixed point. Although our improved numerical computations support this
analytical result for $d=7$, for $d=6$ the approach to the f.p.\ is too close
to resolve this question numerically.

The limiting solution for $d=5$ described in $3.$ is related to a new type of
fixed point described in Section~\ref{sectp}.
The resulting structure is comparable to
the limit of infinite Higgs mass for magnetic monopoles \cite{BFM1}.

As to $4.$ we show that multi-node solutions exist, but only for
$d$ between 4 and 5, i.e.\ not for the integer dimensions studied in
\cite{BCT,BCHT}.

In addition to clarifying the points $1.-4.$ above, we have carried our
conclusions further, to verify that these qualitative features repeat
themselves for the general $p$-$q$-models. Particularly the models with $q=p+1$
show a certain degree of periodicity modulo $4p+1$.

\section{The reduced one dimensional theory}\label{sectred}
The Lagrangean of the Yang--Mills hierarchy in $d$ spacetime dimensions
introduced in \cite{BCT} including the generic $p$-th member, is
\be\label{lagr}
{\cal L}_{YM}=\sum_p \frac{\tau_p}{2\cdot (2p)!}\ \tr\,F(2p)^2 \;,
\ee
where the $2p$ form $F(2p)$ is the totally antisymmetrised product of
the YM field strength $F(2)$.
For the metric the static spherically symmetric ansatz
with $d-1$ space-like dimensions is
\be
\label{general}
ds^2=ds^2_2-r^2d\Omega^2_{d-2}\;,
\ee
where $ds^2_2$ is the metric on the two dimensional orbit space factorising out
the action of the rotation group and $d\Omega^2_{d-2}$ the invariant line
element of $S^{d-2}$. The 2d metric can always be brought to the diagonal form
\be\label{2line}
ds_2^2=e^{2\nu}dt^2-e^{2\lambda}dR^2\;.
\ee
Considering only time independent solutions we naturally take $t$ to be the
Killing time, while the radial coordinate $R$ remains arbitrary.
Plugging this ansatz into the standard Einstein-Hilbert action results
in the reduced one dimensional Lagrangean
\be
\label{grlag}
L_G=-\frac{1}{2G}
\e^{\nu+\la}\,r^{d-4}\Biggl[d-3+\e^{-2\la}\Bigl(\nu'(r^2)'+
    (d-3)(r')^2\Bigr)\Biggr]\;,
\ee
where $G$ denotes Newton's constant.
The static spherically symmetric Ansatz for the $SO_{\pm}(d)$ resp.
$SO_{\pm}(d-1)$ in even resp. odd spacetime dimensional YM field is
\be
\label{YMsph}
A_0=0\;,
\quad
A_i=\left(\frac{1-W}{r}\right)\Sigma_{ij}^{(\pm)}\hat x_j\;,
\quad
\Sigma_{ij}^{(\pm)}=-\frac{1}{4}\left(\frac{1\pm\Gamma_{ch}}{2}\right)
[\Gamma_i, \Gamma_j]\;,
\ee
where $\Gamma_{ch}$ is the chiral matrix in the appropriate dimensions.

The resulting reduced one-dimensional Lagrangean is
\be
L_{\rm YM}=\sum_p r^{d-4}\,V^{p-1}
\left[a_p\e^{\nu-\la}(W')^2+b_p\e^{\nu+\la}r^2V\right]\;,
\ee
with $V=(W^2-1)^2/r^4$ and coefficients
\be
a_p=\tau_p\frac{(d-2)!}{2(2p-1)!(d-2p-1)!}=\frac{1}{2}\hat{\tau_p}\;,
\qquad{\rm and}\qquad
b_p=\frac{d-2p-1}{2p}a_p\;.
\ee
From the form of these cofficients one can see that if $d$ is an integer
it must be larger than $2p$ for the term with $F(2p)$ to contribute.
Thus e.g.\ for $d=4$ only the $F(2)$ term is present. Therefore we will
restrict our analysis to values of $d>2P$, where $P$ is the maximal
value of $p$ in the sum.

An infinitesimal scale transformation of the
action $S=\int dR(L_G+L_{YM})$ yields a virial theorem. While the
gravitational part picks up a factor $d-3$ the $F(2p)$ terms get a factor
$d-4p-1$. As a consequence non-trivial flat solutions exist only, if
the conditions $d-4P-1<0$ and simultaneously $d-4p-1>0$ for the first
non-vanishing term in the YM part are fulfilled.
On the other hand, if $d-4p-1<0$ for the first
non-vanishing term then the gravitational part is required to obtain
non-trivial solutions.

Varying the action with respect to the variables $\nu$, $r$ and $W$
yields three second order ordinary differential equations.
Introducing independent variables $N$, $\ka$ and $U$ for the first derivatives
\begin{equation}\label{first}
N\equiv e^{-\lambda}r'\;,
\quad
\kappa\equiv re^{-\lambda}\nu'+N\;,
\quad
U\equiv e^{-\lambda}W'\;,
\end{equation}
the equations of motion become
\bea\label{feq}
re^{-\lambda}N'&=&(\ka-N)N-2G\sum_p a_p V^{p-1}U^2\;,
\\
re^{-\lambda}\ka'&=&-\frac12(d-3)(d-6)-\ka^2+\frac12(d-4)(d-5)N^2\\
&&-G\sum_p V^{p-1}\Bigl(a_p(d-4p-2)U^2-b_p(d+4p-8)T^2\Bigr)\;,
\\
re^{-\lambda}U'&=&\frac{1}{\sum_{p}a_p\,V^{p-1}}\Biggl[
\sum_p V^{p-1}\Bigl((4p+1-d)N-\ka\Bigr)a_p U\\
 &&+2WT\sum_p V^{p-1}\Bigl(pb_p-(p-1)a_p\frac{U^2}{T^2}\Bigr)\Biggr]\;,
\eea
where we have introduced the shorthand $T=(W^2-1)/r$.
The variation with respect to $\lambda$ yields the algebraic constraint
for the first derivatives
\be
\ka N-\frac12(d-3)+\frac12(d-5)N^2
  -G\sum_p V^{p-1}\Bigl(a_p U^2-b_p T^2\Bigr)=0\;.
\label{constraint}
\ee
Subsequently we shall restrict ourselves to models with only two terms
$(p1,p2)$ in the sum over $p$ denoting them simply as $p$ and $q$. The
simplest choice is $p=1$ and $q=2$, the choice made in \cite{BCT,BCHT}.
Suitably rescaling $r$ and the total action the model depends only on
the dimensionless parameter
$\alpha^{4(q-p)}=G^{2(q-p)}\hat{\tau_p}^{2q-1}\hat{\tau_q}^{1-2p}$.
Taking more terms in the sum does not seem to change the overall picture,
but makes the numerical as well as the analytical analysis much more
cumbersome.

\section{The $F(2p)$ model}\label{sectp}
Before we turn to the $F(2p)+F(2q)$ model we consider the model with
only one $F(2p)$ term, generalising the system studied by Bartnik and
McKinnon \cite{BM}.
In this case the parameters $G$ and $\hat\tau_p$ can be completely absorbed
by a rescaling of $r$ and the total action.
We still have the freedom of choosing a suitable radial coordinate.
As in \cite{BFM} we impose the condition $e^\lambda=r$, call the
corresponding radial coordinate $\tau$ and denote $\tau$ derivatives by a dot.
Thus we obtain from Eqs.~(\ref{feq})
\bea\label{fpeq}
\dot r&=&rN\;,
\\
\dot W&=&rU\;,
\\
\dot N&=&(\ka-N)N-V^{p-1}U^2\;,
\\
\dot\ka&=&1-\ka^2+\frac{(d-4)(d-5)}{2}(N^2-1)\nonumber\\
     && +\frac{V^{p-1}}{2}\Bigl((4p+2-d)U^2
        +\frac{(d-2p-1)(d+4p-8)}{2p}T^2\Bigr)\;,
\\
\dot U&=&\Bigl((4p+1-d)N-\ka\Bigr)U
     +WT\Bigl(d-2p-1-2(p-1)\frac{U^2}{T^2}\Bigr)\;,
\eea
and the constraints
\bea\label{constr}
&&W^2-1-rT=0\;,
\\
&&2\ka N+3-d+(d-5)N^2-V^{p-1}\Bigl(U^2 -\frac{d-2p-1}{2p}T^2\Bigr)=0 \;.
\eea

It turns out to be convenient to introduce new variables
$y=V^{p-1}T^2$ and $z=U/T$ and rewrite the equations as
\bea\label{feqmoda}
\dot r&=&rN\;,
\\
\dot W&=&rTz\;,
\\
\dot T&=&(2Wz-N)T\;,
\\
\refstepcounter{equation}\label{feqmodb}
\dot N&=&(\ka-N)N-yz^2\;,
\\
\dot\ka&=&1-\ka^2+\frac{(d-4)(d-5)}{2}(N^2-1)\nonumber\\
     && +\Bigl((4p+2-d)z^2
        +\frac{(d-2p-1)(d+4p-8)}{2p}\Bigr)\frac{y}{2}\;,
\\
\dot y&=&\Bigl(4pWz-(4p-2)N\Bigr)y\;,
\\
\dot z&=&(d-2p-1)W+\Bigl((4p+2-d)N-\ka\Bigr)z-2pWz^2\;,
\eea
and the constraints
\bea\label{conmod}
&&2\ka N+3-d+(d-5)N^2-\Bigl(z^2 -\frac{d-2p-1}{2p}\Bigr)y=0\;,
\\
&&W^2-1-rT=0\;,
\\
&&T^{2p}-r^{2p-2}y=0\;.
\eea

\section{Fixed Points of the $F(2p)$ model}\label{sectfp}
\subsection{$r=0$ and $r=\infty$}\label{fpr0}
Looking for globally regular solutions existing for $0\leq r<\infty$
we have to cope with the singularities of Eqs.~(\ref{fpeq}) at $r=0$
and $r=\infty$. Since for regular solutions the space time is locally flat
at $r=0$ and $r=\infty$, we have to require $N=\ka=1$ there.
This implies that the radial variable $\tau$ behaves locally like $\ln r$
and thus $\tau\to\mp\infty$ for $r\to 0$ resp.\ $r\to\infty$.
With a suitable choice of new dependent
variables both singular points become hyperbolic fixed points (f.p.s).
From this viewpoint regular solutions correspond to the
`stable manifold' of the respective fixed point \cite{Hart}. As in the case of
the BM solutions regularity requires $W=\pm 1$ and $U=0$ at $r=0$ and~$\infty$.
Let us first turn to the point $r=0$. Although a Taylor
series expansion using the `Schwarzschild' coordinate $r$ as an independent
variable yields essentially one arbitrary parameter $b$ from
$W=1-br^2+O(r^4)$ as in the case of the $F(2)$ theory \cite{BM,BFM},
the linearisation at the f.p.\ is slightly more difficult due to the appearance
of the singular expression $V=(W^2-1)^2/r^4$ and the auxiliary variable $z$
in the equations.

For $W=N=\ka=1$ the equation for $z$ has the f.p.s $z=1$ and~$(2p+1-d)/2p$.
It turns out that $z=1$ for solutions regular at $r=0$, while
$z=(2p+1-d)/2p$ yields a singular mode. This suggests the introduction of the
variables
\be\label{desing}
\bar W=\frac{W-1}{r^2}\;,
\quad
\bar z=\frac{z-1}{r}\;,
\quad
\bar N=\frac{N-1}{r}\;,
\quad{\rm and}\quad
\bar\ka=\frac{\ka-1}{r}\;.
\ee
While $\bar z$, $\bar N$, and $\bar\ka$ vanish at $r=0$ the variable
$\bar W$ has the finite limit $-b$, where $b$ is the parameter
of the Taylor expansion. From Eqs.~(\ref{fpeq}) we obtain
\bea\label{rzero}
\dot{\bar W}&=&O(r)\;,
\\
\dot{\bar z}&=&-d\bar z+(4p+1-d)\bar N-\bar\ka+O(r)\;,
\\
\dot{\bar N}&=&-\bar N+\bar\ka+O(r)\;,
\\
\dot{\bar\ka}&=&(d-4)(d-5)\bar N-2\bar\ka+O(r)\;.
\eea
Replacing the $\tau$ derivatives by $r\,d/dr=(1+O(r))d/d\tau$
these equations take the form required by Prop.~1 of \cite{BFM}.
It is straightforward to see that besides
the regular $r^2$ mode for $W$ the linearisation yields a singular one
behaving as $r^{(2p+1-d)/p}$.
The $(N,\ka)$ system yields an eigenvalue $3-d$, and a second one $d-6$
incompatible with the constraint Eq.~(\ref{constr}b).
Thus for $d>3$ the behaviour of $N$ and $\ka$ at $r=0$ is determined through
the non-linear terms in Eqs.~(\ref{rzero}). One finds
\bea\label{nkap}
N&=&1-\frac{(4b^2)^p}{4p}r^2+O(r^4)\;,
\\
\ka&=&1+(4p-3)\frac{(4b^2)^p}{4p}r^2+O(r^4)\;.
\eea

The linearisation for $r\to\infty$ proceeds along similar lines.
One finds that regular solutions behave like
\be\label{defc}
W=\pm (1-cr^{(2p+1-d)/p}+\ldots)\;,
\ee
and there is a singular
$r^2$ mode. Thus the regular and singular modes of $W$ exchange their
role going from $r=0$ to $r=\infty$ as to be expected from
$\tau\to\pm\infty$. Similarly the eigenvalue $3-d$ from the $(N,\ka)$
system now yields a regular mode
\be\label{defm}
N=1-\frac{m}{2r^{d-3}}+\ldots\;,
\ee
and solutions regular at $r=\infty$ are described by two free parameters,
$c$ and the `mass' $m$.

\subsection{$W=U=0$}\label{fpw0}
Besides the f.p.s for $r=0$ and $r=\infty$ there is also a f.p.\
for finite $r=r_0$ with $W=U=N=0$ and $\ka=\ka_0$.
This f.p.\ is characteristic for the extremal Reissner-Nordstr{\o}m
solution at its horizon
and therefore we call it the RN f.p.%
\footnote{The RN field configuration
arising in the $d=4$ EYM theory is the infinite node limit of the regular
solutions, in the external region the function $W(r)$ vanishing everywhere.
In that case the $SO_{\pm}(4)=SU_{\pm}(2)$ gauge group breaks down to
$SO(2)=U(1)$, namely this field configuration is an Abelian embedding. In the
present cases, the $W(r)=0$ configuration of a $SO_{\pm}(N)$ gauge field with
$N\ge 5$, describes a $SO(N-1)$ gauge field, namely a non-Abelian embedding.}

From the constraints Eqs.~(\ref{constr}) we find
\be
r_0^{2-4p}=\frac{2p}{(d-2p-1)(d-3)}\;,
\qquad{\rm and}\qquad
\ka_0^2=(2p-1)(d-3)\;.
\ee
Putting $r=r_0(1+\delta r)$, $rU=\bar U$ and $\ka=\ka_0+\delta\ka$
and keeping only linear terms we obtain from Eqs.~(\ref{feq}) the system
(eliminating $\delta r$ via Eq.~(\ref{constr}))
\bea\label{RNfp}
\dot{N}&=&\ka_0 N\;,
\\
\dot{\delta \ka}&=&-2\ka_0\delta\ka-(d+4p-8)\ka_0 N\;,
\\
\dot{W}&=&\bar U\;,
\\
\dot{\bar U}&=&-(d-2p-1)W-\ka_0\bar U\;.
\eea
The eigenvalues of these linear equations are $\ka_0,-2\ka_0$ for the
$(N,\ka)$
system and
\be\label{eigeng}
\lambda_{\pm}=\frac{1}{2}\Bigl(-\ka_0\pm\sqrt{(2p-5)d+2p+7}\Bigr)\;,
\ee
for the $(W,\bar U)$ system.
The eigenvalues $\lambda_{\pm}$ are always real for $p>2$, while for
$p=1$ and $p=2$ they are real only if $d\leq 3$ resp.\ $d\leq 11$.

There is a similar f.p.\ with $N=\ka=1$ and $W=U=0$ already present for
the flat $F(2p)$ model. Since then
$\tau\approx\ln{r}$ this f.p.\ requires $r\to 0$ or $r\to\infty$.
The corresponding eigenvalues $\lambda_{\pm}$ are
\be\label{eigenf}
\lambda_{\pm}=\frac{1}{2}\Bigl(4p+1-d\pm\sqrt{(d-4p-3)^2-8p-4}\Bigr)\;.
\ee
These eigenvalues are real if $d$ lies outside the interval
$[4p+3-2\sqrt{2p+1},4p+3+2\sqrt{2p+1}]$.
For $p=2$ this interval is $[6.528,15.472]$ and for $p=3$ it is
$[9.709,20.292]$.
As long as the eigenvalues have non-vanishing imaginary part the solutions
coming close to the f.p.\ are oscillating. As a consequence we might expect to
find regular solutions with any number of zeros.
However, for $d>4p+1$ the real part of $\lambda_{\pm}$ becomes negative and
thus the solutions coming close to the f.p.\ tend to zero for $r\to\infty$
and $W$ cannot reach $\pm 1$. In order to analyse these
oscillating solutions beyond the linear approximation we consider the `Lyapunov
Function'
\be\label{Lyap}
Z=\Bigl(z^2-\frac{d-2p-1}{2p}\Bigr)r^{4p-2}y
 =(W^2-1)^{2p-2}\dot W^2-\frac{d-2p-1}{2p}(W^2-1)^{2p}\;,
\ee
which is a measure of the amplitude of oscillations (as long as $|W|<1$),
and satisfies
\be\label{dLyap}
\dot Z=2\Bigl((4p+2-d)N-\ka\Bigr)(W^2-1)^{2p-2}\dot W^2\;.
\ee
Thus for asymptotically flat solutions with $N\to1$, $\ka\to1$ as
$r\to\infty$, the amplitude of oscillations decreases for $d>4p+1$,
increases for $d<4p+1$, and has a finite limit for $d=4p+1$.
That means that regular solutions with many zeros cease to exist for
$d\ge4p+1$. However, since $4p+3-2\sqrt{2p+1}<4p+1$ there is always a finite $d$
interval, where regular solutions with any number of zeros are expected to
exist.

\subsection{Conical singularity}\label{conical}
For $p>1$ there is another f.p., which turns out to be relevant for the
limit $b\to\infty$ in certain dimensions (e.g.\ $d=5$ for the $F(2)+F(4)$
model).
It is characterised by $W=1$ and nontrivial ($\neq 1$) finite values of $N$,
$\ka$ and $z$.
Putting the r.h.s.\ of Eqs.~(\ref{feqmodb}) to zero yields the f.p.\ equations
\bea\label{fpeqs}
0&=&\ka N-N^2-yz^2\;,
\\
0&=&1-\ka^2+\frac{(d-4)(d-5)}{2}(N^2-1)+\frac{(4p+2-d)}{2}yz^2\nonumber\\
  &&  +\frac{(d-2p-1)(d+4p-8)}{4p}y\;,
\\
0&=&\Bigl(4pz-(4p-2)N\Bigr)y\;,
\\
0&=&(4p+2-d)Nz-\ka z+d-2p-1-2pz^2\;,
\eea
leads to the solution
\bea\label{fpsln}
N_0^2&=&\frac
{16p^{3}-8dp^{2}+2pd^{2}-8pd+14p+d-3\pm\sqrt{Q}}{2(4p^2-4p+1)(2p-1)}\;,
\\
\ka_0&=&(2p+3-d)N_0+\frac{2p(d-2p-1)}{(2p-1)N_0}\;,
\\
y_0&=&\frac{4p^2}{(2p-1)^2}\Bigl(\frac{\ka_0}{N_0}-1\Bigr)\;,
\\
z_0&=&\frac{2p-1}{2p}N_0\;,
\eea
where
\be
Q = (2 p d - 6 p + 1) (d - 3)
    (2 p d^2 - 16 p^2 d - 4 p d + d + 32 p^3 + 10 p - 3)\;.
\ee
The zeros of the polynomial $Q$ are at $d=(6p-1)/2p$, $3$, and
$(16p^2+4p-1\pm\sqrt{128p^3-96p^2+16p+1})/4p$. In order to obtain
a real value for $N_0$ in the admissible region the dimension $d$ has to be
restricted to the interval
$3\leq d\leq d_p$ with
\be\label{defdp}
d_p=\frac{16p^2+4p-1-\sqrt{128p^3-96p^2+16p+1}}{4p}\;,
\ee
or it has to be larger than
$(16p^2+4p-1+\sqrt{128p^3-96p^2+16p+1})/4p$.
The numerical values of $d_p$ for $p=2$, $3$, and $4$ are $\approx 5.63$, $8.63$, and
$11.81$.

The fact that $N$ has a finite non-vanishing limit $N_0<1$ implies that $r\to0$ for
$\tau\to-\infty$. As a consequence of $N_0\ne1$ the gravitational field has a `conical'
singularity at $r=0$ with a solid angle deficit and the
YM field has a weak singularity
\be\label{sing0}
W=1-\frac{1}{2}y_0^{1/2p}r^{2-1/p}+O(r^{2+1/p})\;,
\ee
as for the massive EYM model (`Proca theory') studied in \cite{BFM1}.

In order to linearise the Eqs.~(\ref{feqmodb}) at this `conical' f.p.\ we put
$N=N_0+\delta N$, $\ka=\ka_0+\delta \ka$, $y=y_0+\delta y$, $z=z_0+\delta z$
and keep only linear terms in $\delta N$ etc. Eliminating $\delta\ka$ via the
constraint Eq.~(\ref{conmod}a) we obtain
\bea\label{lin}
\dot{\delta N}&=&(3-d)N_0\delta N
                 -\Bigl(\frac{z_0^2}{2}+\frac{d-2p-1}{4p}\Bigr)\delta y
                 -y_0z_0\delta z\;,
\\
\dot{\delta y}&=&-2(2p-1)y_0\delta N+4py_0\delta z\;,
\\
\dot{\delta z}&=&\frac{2p-1}{2p}\Bigl((4p-3)N_0+\ka_0\Bigr)\delta N
        +\frac{2p-1}{4p}\Bigl(\frac{d-2p-1}{2p}-z_0^2\Bigr)\delta y\nonumber\\
        && +\Bigl((4-d)N_0-\ka_0-\frac{2p-1}{2p}y_0z_0\Bigr)\delta z\;.
\eea
Due to the complicated structure of this linear system its eigenvalues
can only be determined numerically. For small $p$ one finds that
there is one pair of complex conjugate eigenvalues if $d$ is
near $d=2p$, but there are three real ones if $d$ is larger than some
value $d^{(r)}_p$. For $p=2$ resp.\
$p=3$ one finds $d^{(r)}_2\approx 5.55$ and $d^{(r)}_3\approx 8.0$.
For $p\approx 5.2$ the value $d^{(r)}_p$ becomes equal to $2p$ and all
three eigenvalues are real for all $d\geq 2p$.
For $d=d_p$ one of the three negative modes becomes a zero mode, because
the two different solutions for $N_0^2$ degenerate.
As long as there are complex eigenvalues the solutions coming close to the
f.p.\ oscillate.

Since the real parts of all three eigenvalues are negative for $d<d_p$ the
f.p.\ is always repulsive for $\tau\to-\infty$. Thus there are no free
parameters available that would be necessary to construct global solutions
regular at $r=\infty$ with the behaviour (\ref{sing0}) near $r=0$.

\section{$F(2p)+F(2q)$ model; $q>p$}\label{sectpq}
Next we study the model with with the two terms $F(2p)+F(2q)$. As mentioned
above we require $d>2q$  (and hence $d>2p+2$) and $d<4q+1$.
After a suitable rescaling this model depends on a single dimensionless parameter
$\alpha$.

From Eqs.~(\ref{feq}) we get
\bea
\dot r&=&rN\;,
\\
\dot W&=&rU\;,
\\
\refstepcounter{equation}\label{pqeq}
\dot N&=&(\ka-N)N-\alpha^2V^{p-1}(1+V_{pq})U^2\;,
\\
\dot\ka&=&1-\ka^2+\frac{(d-4)(d-5)}{2}(N^2-1)\nonumber\\
      && +\frac{\alpha^2}{2}V^{p-1}\Biggl[\Bigl(4p+2-d
    +(4q+2-d)V_{pq}\Bigr)U^2
    +\Bigl(\frac{(4p-8+d)(d-2p-1)}{2p}\nonumber\\
    &&+\frac{(4q-8+d)(d-2q-1)}{2q}V_{pq}\Bigr)T^2\Biggr]\;,
\\
\dot U&=&\frac{1}{1+V_{pq}}\Biggl[\Bigl((4p+1-d)N-\ka\Bigr)U
     +\Bigl((4q+1-d)N-\ka\Bigr)V_{pq}U\nonumber\\
    && +WT\Bigl(d-2p-1+(d-2q-1)V_{pq}
        -2(p-1+(q-1)V_{pq})\frac{U^2}{T^2}\Bigr)\Biggr]\;,
\eea
and the constraint
\begin{eqnarray}
2\ka N+3-d+(d-5)N^2&=&\alpha^2V^{p-1}\Biggl[\Bigl(1+V_{pq})U^2
  -\Bigl(\frac{d-2p-1}{2p}\nonumber\\
&&+\frac{(d-2q-1)}{2q}V_{pq}\Bigr)T^2\Biggr]\;,
\end{eqnarray}
with $V_{pq}=V^{q-p}$.

Again it turns out to be useful to introduce the variables
$y=V^{p-1}T^2$ and $z=U/T$ and rewrite the equations as
\bea\label{pqzeqa}
\dot r&=&rN\;,
\\
\dot W&=&rTz\;,
\\
\refstepcounter{equation}\label{pqzeqb}
\dot N&=&(\ka-N)N-\alpha^2(1+V_{pq})yz^2\;,
\\
\dot\ka&=&1-\ka^2+\frac{(d-4)(d-5)}{2}(N^2-1)
       +\frac{\alpha^2}{2}\Biggl[\Bigl(4p+2-d+
     (4q+2-d)V_{pq}\Bigr)z^2\nonumber\\
     && +\Bigl(\frac{(4p-8+d)(d-2p-1)}{2p}+
            \frac{(4q-8+d)(d-2q-1)}{2q}V_{pq}\Bigr)\Biggr]y\;,
\\
\dot y&=&4p\Bigl(Wz-\frac{2p-1}{2p}N\Bigr)y\;,
\\
\dot z&=&\frac{1}{1+V_{pq}}\Biggl[\Bigl((4p+2-d)N-\ka\Bigr)z
     +\Bigl((4q+2-d)N-\ka\Bigr)V_{pq}z\nonumber\\
    && +W\Bigl(d-2p-1+(d-2q-1)V_{pq}
                   -2(p+qV_{pq})z^2\Bigr)\Biggr]\;,
\eea
with the constraints
\bea
2\ka N+3-d+(d-5)N^2&=&\alpha^2\Biggl[\Bigl(1+V_{pq})z^2\nonumber\\
 && -\Bigl(\frac{d-2p-1}{2p}+\frac{(d-2q-1)}{2q}V_{pq}\Bigr)\Biggr]y\;,
\\
W^2-1-rT&=&0\;,
\\
T^{2p}-r^{2p-2}y&=&0\;.
\eea

The behaviour of solutions regular at $r=0$ is now given by the expansions
\bea\label{nkpq}
N&=&1-\alpha^2\Bigl(\frac{(4b^2)^p}{4p}+\frac{(4b^2)^q}{4q}\Bigr)r^2+O(r^4)\;,
\\
\ka&=&1+\alpha^2\Bigl((4p-3)\frac{(4b^2)^p}{4p}+(4q-3)\frac{(4b^2)^q}{4q}\Bigr)r^2
    +O(r^4)\;.
\eea

The $F(2p)$ and $F(2q)$ terms in Eqs.~(\ref{nkpq}) scale differently.
In the limit  $b\to\infty$ the $F(2p)$ term decouples and one obtains the
BM type solutions of the $F(2q)$ model after the rescaling
$b\to b\alpha^{\frac{2}{2q-1}}$ and letting $\alpha\to 0$.
Likewise for $b\to 0$ and $d<4p+1$ one obtains the BM type solutions of the
$F(2p)$ model rescaling $b\to b\alpha^{\frac{2}{2p-1}}$ and letting now
$\alpha\to\infty$.

From the virial theorem we know that for $4q+1>d>4p+1$ there is a flat
solution, obtainable in the limit $\alpha\to 0$;
the corresponding parameter $b_{\rm flat}$ diverges as $d\to 4q+1$ and
vanishes in the limit $d\to 4p+1$.

For $d=4p+1$ and $b\to0$ the $F(2q)$ and gravity terms both decouple and one
obtains a rescaled flat $F(2p)$ instanton. Linearising the equations for a
perturbation of the instanton background yields a finite limit for
$b^{4p-2q-1}\alpha^2$; thus $\alpha$ vanishes as $b\to0$ if $4p<2q+1$ or
diverges if $4p>2q+1$.

It is easy to see that for $r\to\infty$ the terms coming from $F(2q)$
are subleading and hence the asymptotic behaviour is that of the $F(2p)$
model.

Also the other f.p.s besides $r=0$ and $r=\infty$
discussed for the $F(2p)$ model have counterparts here.
Of particular relevance
for the application to limiting solutions is the RN f.p.\
with $W=U=N=0$ and finite values of $r$ and $\ka$
determined by the equations
\bea\label{RN24}
0&=&3-d+\frac{\alpha^2}{2r_0^{4p-2}}\Bigl(\frac{d-2p-1}{p}
       +\frac{d-2q-1}{2qr_0^{4q-4p}}\Bigr)\;,
\\
\ka_0^2&=&1-\frac{(d-4)(d-5)}{2}+\frac{\alpha^2}{4r_0^{4p-2}}
      \Bigl(\frac{(4p-8+d)(d-2p-1)}{p}\nonumber\\
        && +\frac{(4q-8+d)(d-2q-1}{2qr_0^{4q-4p}}\Bigr)\;.
\eea
Except for $d=2p+1$ or $d=2q+1$ the resulting equation for $r_0$ can only
be solved numerically.

Similarly there is the f.p.\ with $W=U=0$, $N=\ka=1$ and $r\to\infty$
already present in the flat $p$-$q$-model.
However, for $r\to\infty$ only the $F(2p)$ term survives so we
can use the corresponding results of the previous section.

What remains is the conical f.p.\ of Eqs.~(\ref{fpeqs}). It turns out to
become relevant for the limiting solutions as $b\to\infty$ in the interval
$2q<d<d_q$ (e.g.\ $d=5$ for the 1-2-model). The numerical analysis shows
that for large values of $b$ the solutions regular at $r=0$ come very close
to the f.p.\ (of the $F(2q)$ model) within a very short $r$ interval of
length $r_0=1/b^q$. They consist of an `interior' part between $r=0$ and
$r\approx\sqrt{r_0}$ and an `exterior' one extending from
$r\approx\sqrt{r_0}$ to $r=\infty$ (see Figs.~\ref{r1p1q2d5en}
and~\ref{r1p1q2d5ey}).
\mkfig{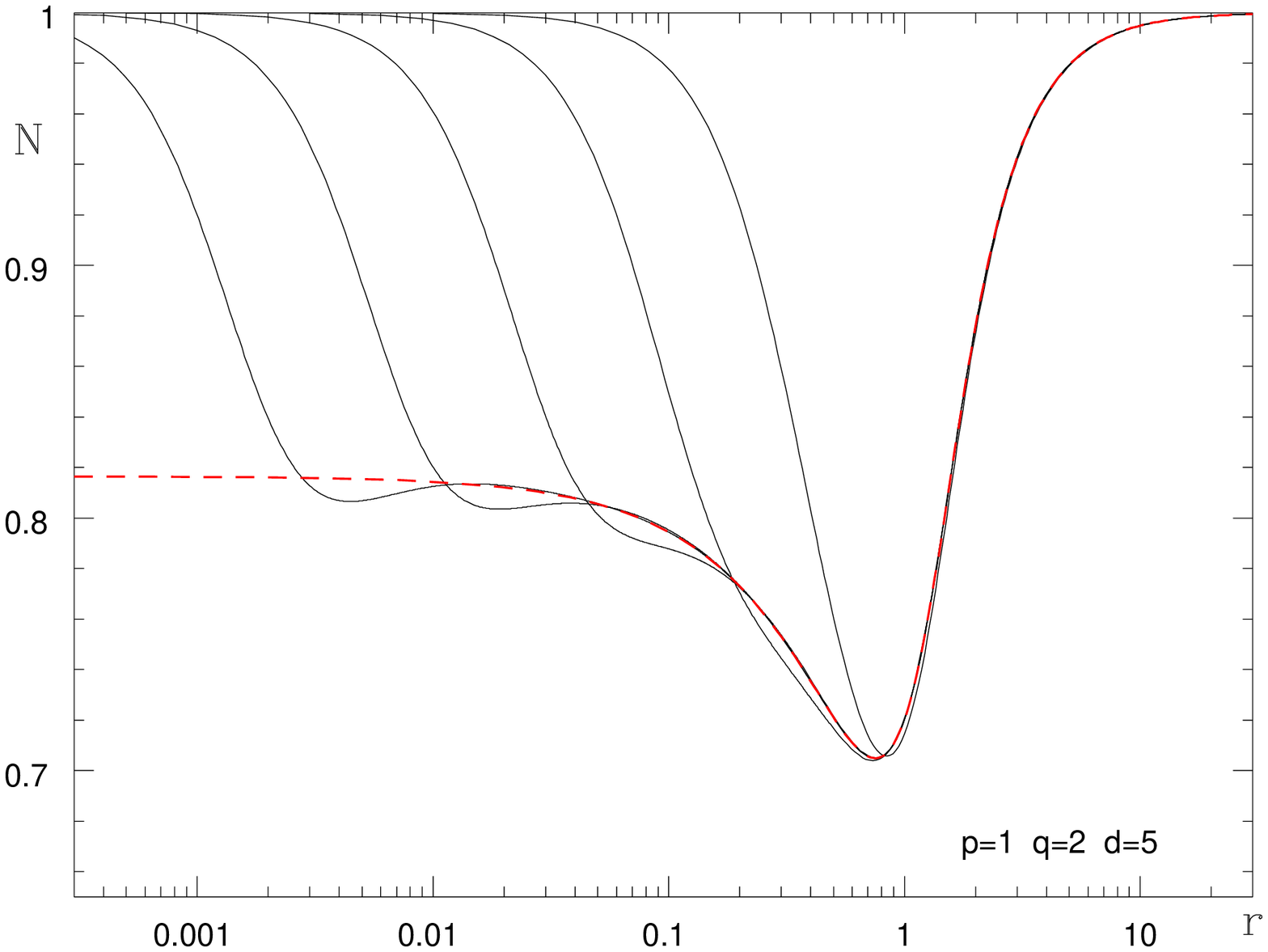}{0.9}{0.9}{45 178 561 566}
{$N(r)$ of the 1-2-model for $d=5$ and increasing values of $b$}
\mkfig{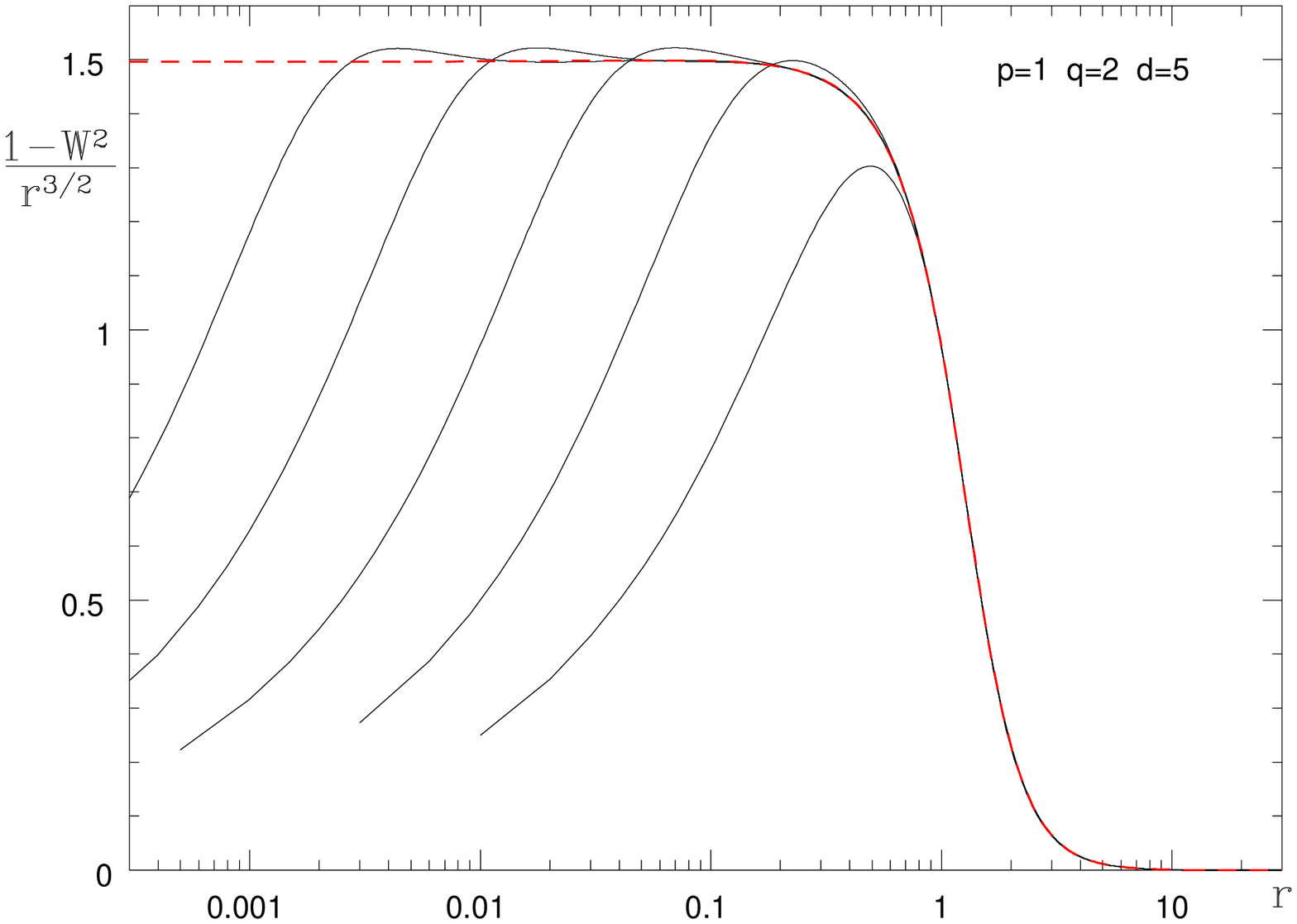}{0.9}{0.9}{45 178 561 566}
{$y^{1/4}=(1-W^2)/r^{3/2}$ of the 1-2-model for $d=5$ and increasing
values of $b$}

The interior part is dominated by the $F(2q)$ terms in Eqs.~(\ref{pqzeqb})
and the expansions Eqs.~(\ref{nkpq}) suggest the introduction of the rescaled
variable $\bar r=r/r_0\equiv rb^q$. In the limit $b\to\infty$ the globally regular
solutions split into an interior part extending from the regular origin
$\bar r=0$ to the conical f.p.\ at $\bar r=\infty$ and an exterior one
extending from the f.p.\ at $r=0$ to $r=\infty$. The spacetime corresponding
to the interior part of the solution has a conical singularity at $\bar
r=\infty$ with a solid angle deficit much like for global monopoles
\cite{barriola}.

Solutions for sufficiently large $b$ can be obtained by matching a rescaled
interior solution to an exterior one in a domain where the linearisation at
the f.p.\ is appropriate. Although the $F(2p)$ terms are a small
perturbation for the interior solution, this is true for the exterior one
only near the f.p., while the $F(2p)$ term becomes dominant for large $r$.

In order to describe the interior part we use the variable $\bar r=rb^q$ and
introduce $\bar T=b^{q-1}T$, such that
$y=T^{2q}r^{2-2q}=\bar T^{2q}\bar r^{2-2q}$.
For large $b$ we get from $y=O(1)$ the estimate
\be
W-1\approx \frac{rT}{2}\sim \bar r^{\frac{2q-1}{q}}b^{1-2q}\;.
\ee
Putting $Y=(N,y,z)$ we can linearise Eqs.~(\ref{pqzeqb}) at the
f.p.\ with $Y=Y_0+\delta Y_I$ to obtain
\be\label{int}
\bar r\frac{d}{d\bar r}\delta Y_I=L\delta Y_I+O({\delta Y_I}^2)
   +O({\bar r}^{\frac{2q-1}{q}}b^{1-2q})\;.
\ee
$L$ here is the matrix on the r.h.s.\ of Eqs.~(\ref{lin}), with some
rescaling due to the appearance of $\alpha^2$ in Eqs.~(\ref{pqzeqb}).
In the region $1\ll {\bar r}^{1/q}\ll b$ we can neglect the $O(.)$
terms to get
\be
\delta Y_I=K\bar r^{\lambda}\eta=K(br)^\lambda \;,
\ee
with some constant matrix $K$ and constant vector $\eta$ ($\lambda$ denoting
the diagonal matrix of eigenvalues).
In order that the exterior solution hits the f.p.\ for $\tau\to -\infty$
we have to fine-tune three parameters.
Since the solutions regular at $r=\infty$ depend only
on the two parameters $c$ and $m$ introduced in Eqs.~(\ref{defc})
and~(\ref{defm}), we have to tune also $\alpha$.
This determines the `critical' value of $\alpha$ which in fact is a way to
compute it more precisely than as a limit of regular solutions starting at
$r=0$.
We put $\xi=(\alpha,m,c)$ with $\xi_0$ the triple parametrising the
solution running into the f.p.
For $\xi=\xi_0+\delta\xi$ the solution misses the f.p.\ but comes close to
it for $|\delta\xi|\ll 1$. Linearisation at the f.p., valid for $r\ll 1$,
gives now for the exterior solution
\be
\delta Y_E\approx Kr^\lambda M\delta\xi\;,
\ee
with some constant matrix $M$.
Matching the interior with the exterior solution at some fixed $r\ll 1$
using these approximations gives
\be
M\delta\xi=b^{q\lambda}\eta\;.
\ee
If we rescale the $\delta\xi$'s with the smallest real part of the
$\lambda$'s, the rescaled quantities either tend to a limit for
$b\to\infty$ if the eigenvalues are real, or they oscillate about the
limiting value in case they are complex.
This behaviour is well exhibited in Figs.~\ref{r1p1q2d5} and~\ref{r1p1q2d5p56}.
\mkfig{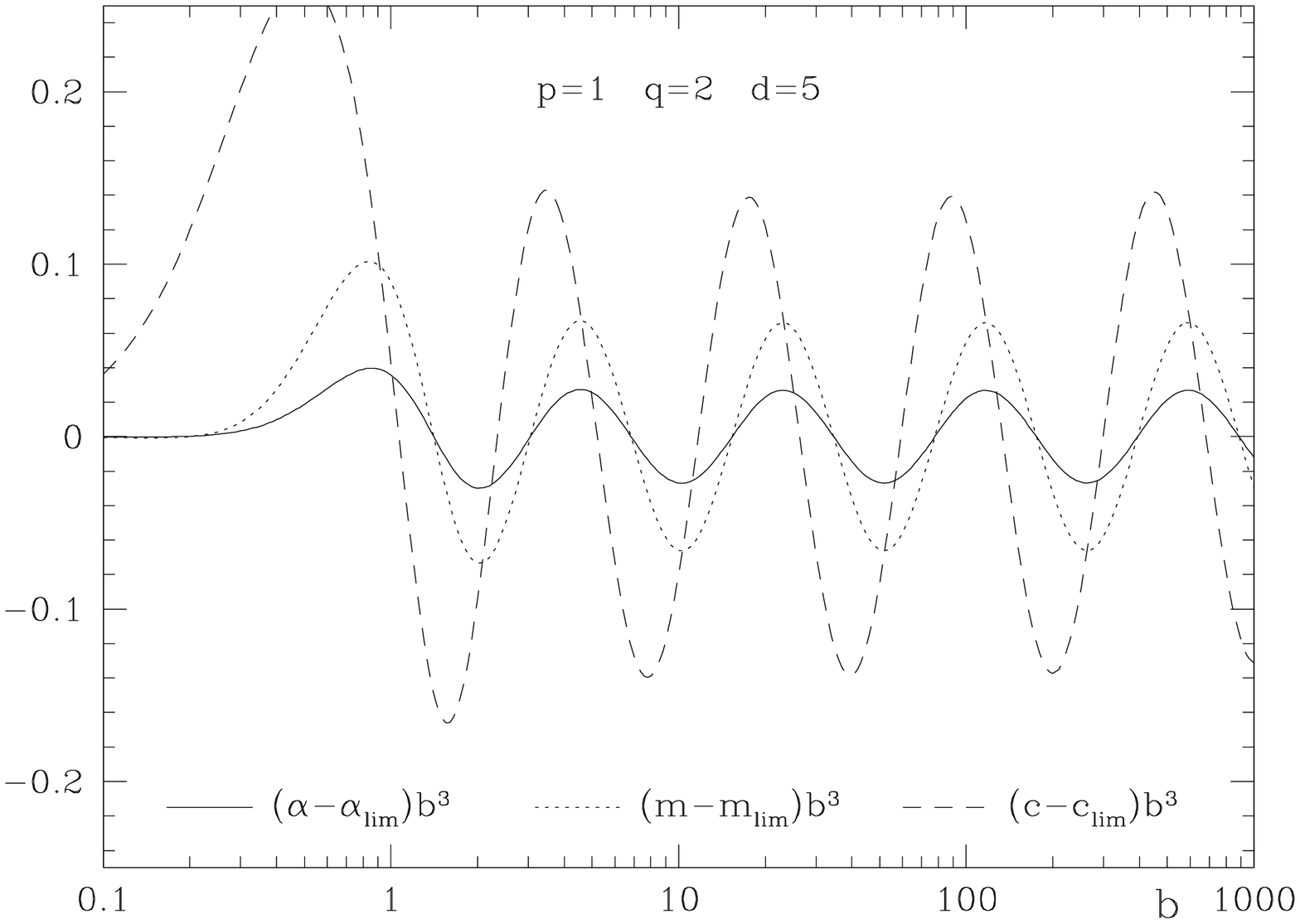}{0.9}{0.9}{45 178 561 566}
{Rescaled parameters of the 1-2-model for $d=5$ and large $b$}
\mkfig{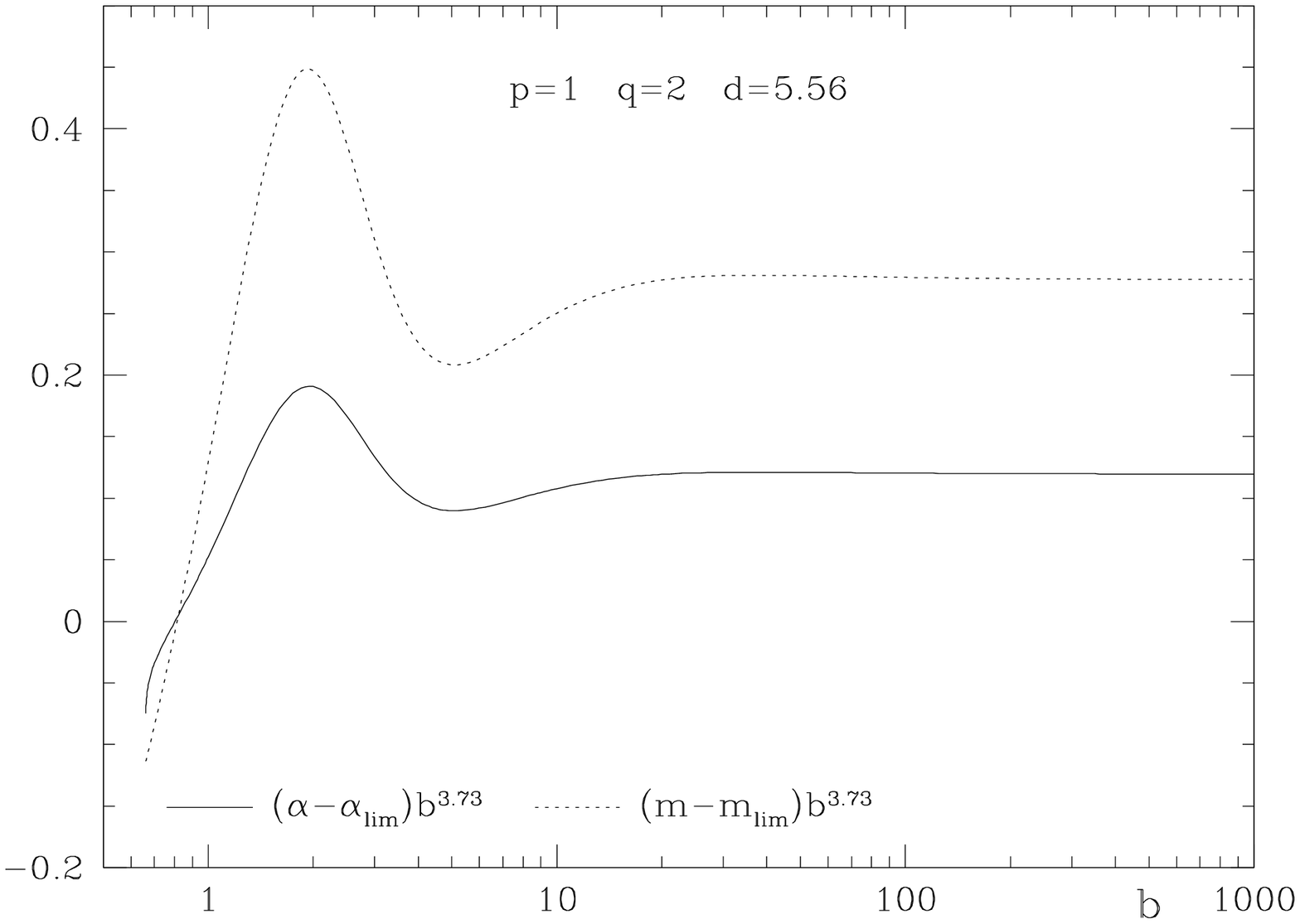}{0.9}{0.9}{45 178 561 566}
{Rescaled parameters of the 1-2-model for $d=5.56$ and large $b$}

\section{Numerical results}\label{sectn}
The numerical analysis of the $F(2)+F(4)$ model \cite{BCT,BCHT}
showed that for $d=5,6,7,$ and $8$ there are 1-parameter families of
globally regular solutions labelled by the parameter $\alpha$.
All these solutions resemble the lowest
BM solution with one zero of the potential $W$. In contrast to the BM case
no solutions with more than one zero were found.
All the families have a maximal
value $\alpha_{\rm max}$  and end at some `critical' value $\alpha_c$,
where the solutions cease to exist. While for
$d=5,6,7$ a critical value $\alpha_c>0$ was obtained, it was suggested that
$\alpha_c=0$ for $d=8$. Our aim is to explain what happens in the limiting
cases, determine the values of
$\alpha_c$ and to analyse the corresponding critical solutions.
As already mentioned above it proves
very helpful for this purpose to consider also non-integer
values of $d$, interpolating between the physically relevant integer ones.
This is made possible by the explicit $d$-dependence of the Eqs.~(\ref{feq}).
In fact we extended our study of the $F(2)+F(4)$
model to the whole interval $4<d<9$ for which globally regular solutions
can be found. We find that for $4<d<5$ analogues of the discrete family of BM
solutions with any number of zeros of $W$ exist,
while for $5\leq d<9$ only solutions with one zero were found.

Let us first discuss the results for $5\leq d< 9$.
Varying the parameter $\alpha$ for a fixed $d$ one obtains a smooth
1-parameter family of solutions
conveniently parametrised by the curves $b(d,\alpha)$, where $b$ is the
parameter characterising the solutions at $r=0$.
These curves
start with finite $b$ at $\alpha=0$ and depending on $d$ either turn back
to $\alpha=0$ after
running through some maximum or end at some finite value $\alpha_c$
(see Fig.~\ref{sp1q2}).
Our aim is to characterise the limiting solutions on both endpoints
of the curve.
\mkfig{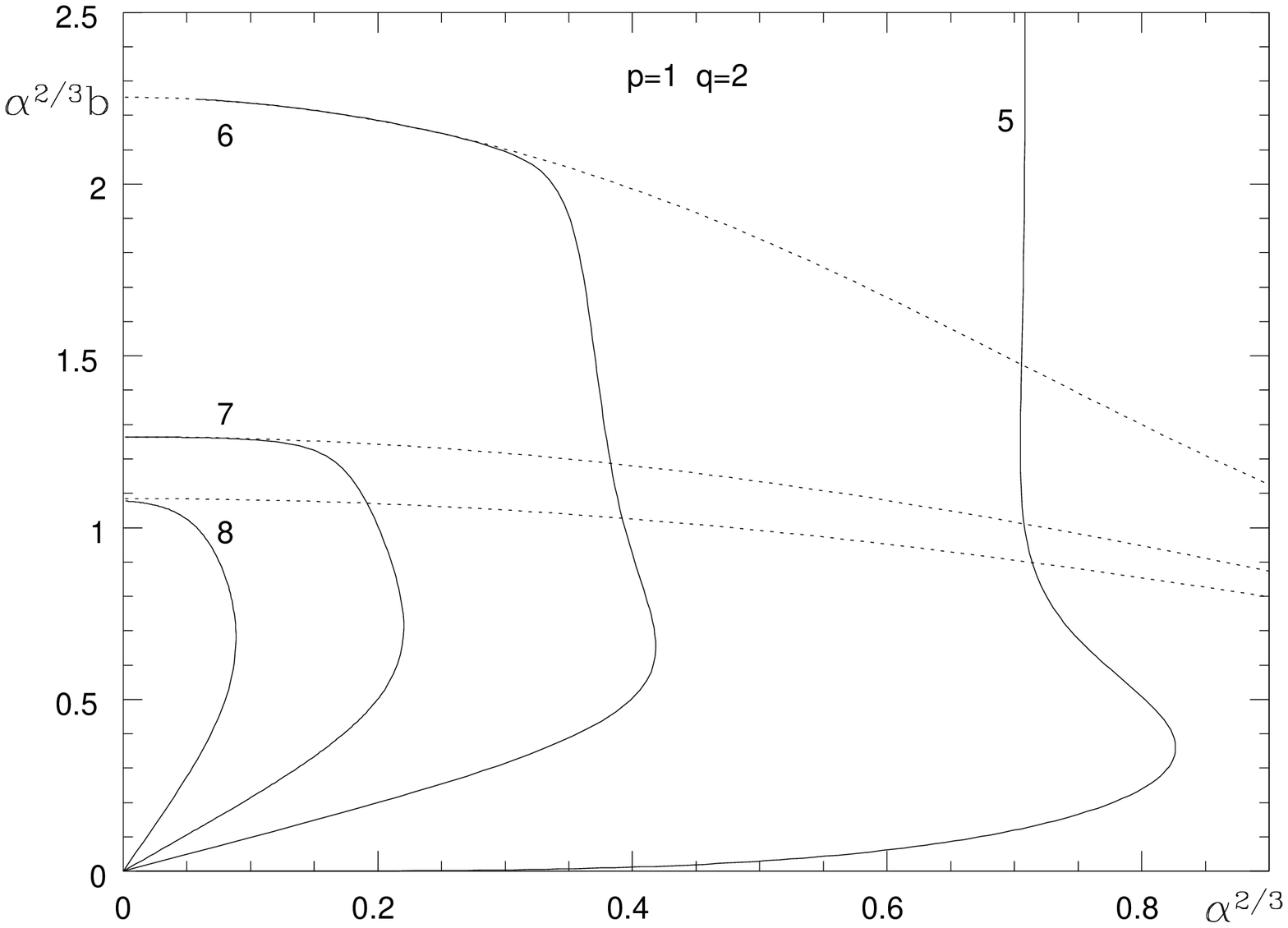}{0.9}{0.9}{45 178 561 566}
{$b$ as a function of $\alpha$ for regular solutions of the 1-2-model
with $d=5$, $6$, $7$, and $8$}

The endpoint $b(d,0)$ is obtained by switching off the gravitational
field taking $\alpha\to 0$. A was discussed in Section~\ref{sectpq}
the corresponding solutions are those of
the flat $F(2)+F(4)$ theory already found some time ago \cite{butc}.
Actually, this holds only for $d>4p+1=5$, while for $d=5$ a slightly
different limit has to be taken switching off also the $F(4)$ term.
More precisely, we have to rescale $r\to r/\alpha$ and simultaneously
$b\to \alpha^2 b$.
The resulting flat space solution is nothing but the 4 dimensional instanton
given by $W=(1-\frac{1}{2}br^2)/(1+\frac{1}{2}br^2)$ considered as a static
soliton of
the $d=5$ theory. Exactly the same function yields `instanton' solutions for
the $F(2p)$ model in $d=4p+1$.

The other endpoints of the curves $b(d,\alpha)$, obtained for large values
of $b$, are more difficult to explain. These are different for
$d=5$~\cite{BCHT} and $d=6,7,8$~\cite{BCT}. As was
already observed in \cite{BCHT} for $d=5$ there seems to be a critical
value $\alpha_c\approx 0.6$ and there are obviously other turning points of
$b(d,\alpha)$ besides the maximum $\alpha_{\rm max}\approx 0.75$.
For large values of $b$ the solutions show a very distinctive
behaviour. $N$ decreases very quickly to some value $N_0\approx 0.82$,
$W$ stays close to one, while $T^2/r$ and $U^2/r$ tend to finite limits.
We claim that the solution comes closer and closer to the conical f.p.\
(\ref{fpsln}) as $b\to\infty$.
In Section~\ref{sectpq} we gave an analytical description of these solutions
leading to an asymptotic formula for the parameters $\alpha$, $c$, and~$m$
for large $b$ and to an independent determination of $\alpha_{c}$.
The value $\alpha_{c}=0.595965$ obtained this way agrees up to this
precision with the value obtained with a regular solution for $b=1000$.
Employing Eq.~(\ref{fpsln}) for $d=5$ we get $N_0=\sqrt{2/3}\approx 0.8165$
well in accordance with our numerical result (see Fig.~\ref{r1p1q2d5en}).
Also the limiting value of $y^{1/4}=(1-W^2)/r^{3/2}$
read of from Fig.~\ref{r1p1q2d5ey}
agrees well with the analytical result
$y_0^{1/4}=2/\sqrt{3\alpha_{c}}\approx 1.496$.
Fig.~\ref{r1p1q2d5} displays the behaviour of
$b^3(\alpha_{c}-\alpha(b))$ etc.\ for large values of $b$,
which is in accordance with our derivation at the end of
Section~\ref{sectpq}. As also mentioned in Section~\ref{sectp}, the f.p.\
ceases to be oscillatory at $d\approx 5.55$, which is in accordance with the
behaviour shown in Fig.~\ref{r1p1q2d5p56} for $d=5.56$.

The f.p.\ exists only for $d\le d_2$ with
$d_2=(71-\sqrt{673})/8\approx 5.632$ given by Eq.~(\ref{defdp}).
Thus this f.p\ can only
be relevant for the limiting solution up to this value, again well in
accordance with our numerical results as shown in Fig.~\ref{r1p1q2}.
\mkfig{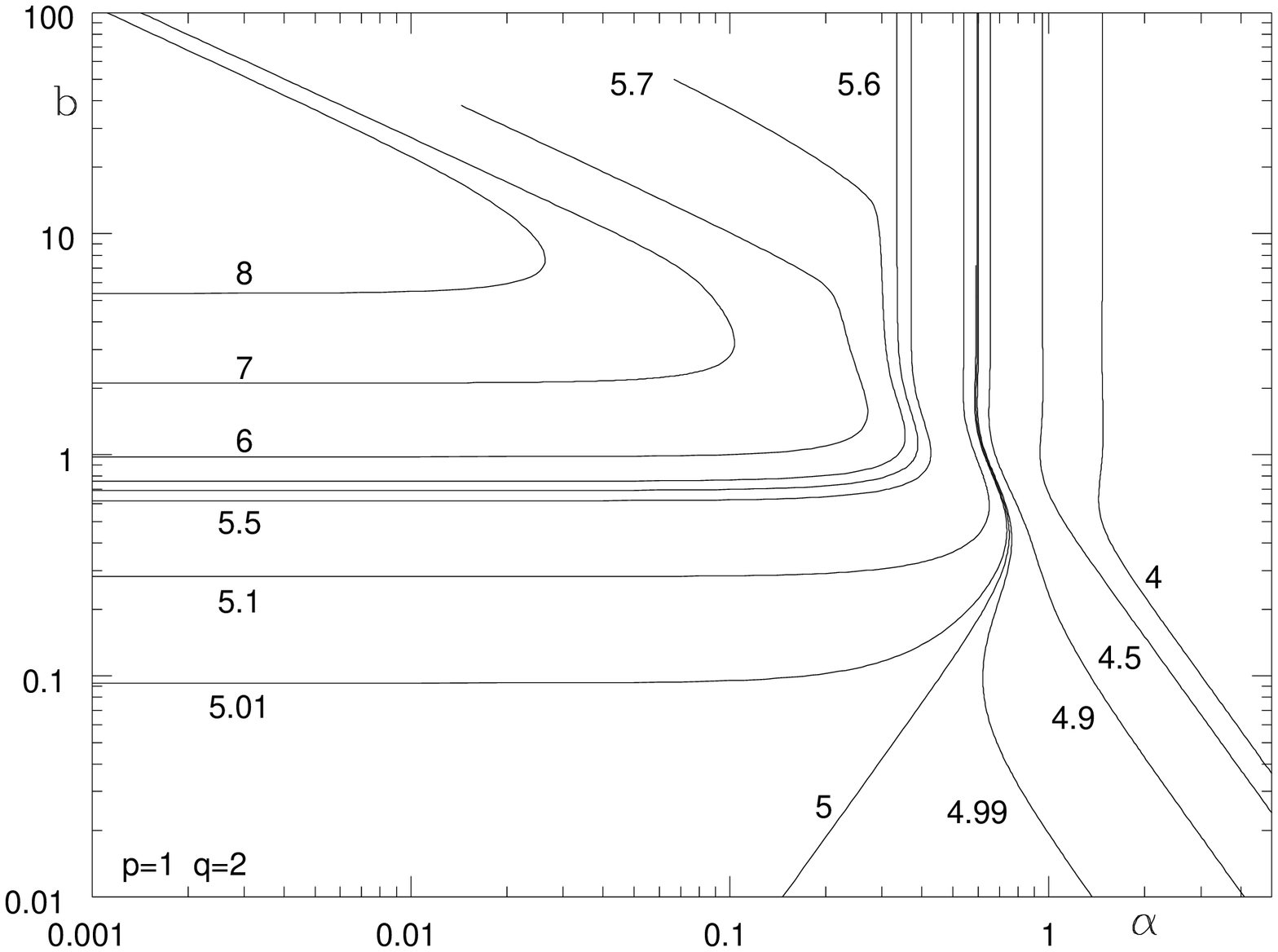}{0.9}{0.9}{45 178 561 566}
{$b$ vs.\ $\alpha$ for regular solutions of the 1-2-model with different
values of $d$}

For $d>d_2$ a different limiting behaviour manifests itself.
As far as the numerical results for $d=6$ are concerned,
the function $N$ seems to develop a double zero at some finite value
of $r$, where the variable $\tau$ increases without limit. This type of
critical behaviour is well known from the limiting solution of the BM family,
when the number of zeros goes to infinity \cite{BFM} and from gravitating
monopoles as described in \cite{BFMmon}. The critical solution runs into
the RN f.p.\ and thus consists of a non-trivial (geodesically complete)
interior solution and an exterior (geodesically incomplete)
one with $W=0$ representing the exterior of the extremal
RN black hole.

Instead of looking for globally regular solutions one may as well look for
solutions starting at $r=0$ and ending at the RN f.p.
Fig.~\ref{sp1q2} shows corresponding curves $b(d,\alpha)$ as dashed
lines. As discussed in Section~\ref{sectpq} we have to rescale
$b\to b\alpha^{2/3}$ in order to obtain the BM solution of the $F(4)$
model. The same rescaling has to be chosen for the solution running into
the RN f.p.
For $d=6$ there seems to be a bifurcation of the globally regular solutions
with these RN ones at $\alpha\approx 0.015$, although numerically it
is clearly
impossible to distinguish a genuine bifurcation from a finite but extremely
small proximity. At any rate, it is rather obvious from
Fig.~\ref{sp1q2} that the
curve $b(d,\alpha)$ for $d=8$ extends all the way to $\alpha=0$ without any
bifurcation with the RN curve, agreeing with the numerical results of
\cite{BCT}.
Although for $d=7$ this is not obvious from Fig.~\ref{sp1q2}
our numerical results clearly exclude a bifurcation. In fact, numerically
we can exclude a bifurcation down to $d\approx 6.8$.
Thus one might expect some
particular value $d=d_b$ for which the bifurcation occurs at $\alpha=0$
and then for $d<d_b$ moves along some curve $\alpha(d)$ down to $d=d_2$.
In the following we shall give a convincing argument,
that actually no such bifurcation happens for any $d_2<d<9$.
This argument is based on an observation made for the BM solutions with
many zeros, i.e.\ close to the limiting solution running into the RN f.p.\
\cite{BFM}.
There, solutions with $n\gg 1$ zeros show three characteristic regions.
Region I extends from
$r=0$ to $r=1$, where the solutions develop more and more zeros of $W$
and come close to the oscillating limiting solution running into the RN
f.p., for which $W$ decays like $e^{-\tau/2}$ for $\tau\to\infty$.
In region II, between $r=1$ and $r=r_n\gg 1$ the
functions $W$ and $U$ first continue to decay, but then they start to grow
like $e^{\tau/2}$, however, still staying very small.
In this interval the metric is well approximated by the extremal RN solution
growing from $N\approx 0$ at $r=1$ to $N\approx 1$ for $r\gg 1$.
Thus the solution is well approximated solving the linearised equation for
$W$ and $U$ in the background of the extremal RN solution.
Finally, in region III between
$r=r_n$ and $r=\infty$ the solutions stay close to the oscillating
flat space solution for $W$ running from $r=\infty$ into the f.p.\ with
$U=W=0$ and $N=\ka=1$. The relevant difference between the RN f.p.\ of the
$(U,W)$ system with $N=0$ and the flat one with $N=1$ manifests
itself in the real part of the eigenvalue being $-1/2$  for $N=0$ and $+1/2$
for $N=1$.

In the present case,
if we assume a bifurcation with the solution running into the RN f.p.\
the same structure with three regions must form.
Now we will argue that this leads to a contradiction.
For $\alpha\neq 0$ the flat f.p.\ with $W=0$ is dominated by the $p=1$
term at large
$r$ and thus is oscillatory in the interval $|d-7|\leq 2\sqrt{3}\approx 3.46$.
Since we are considering a family of solutions with one zero they cannot
come close to this f.p.\ in the considered region with $d\approx 6$.
This means that the curve $b(d,\alpha)$ has to turn back all the way to
$\alpha=0$.
There still remains the possibility that the bifurcation happens at
$\alpha=0$ for some or all $d\leq d_b$. Since $b\to\infty$ as $\alpha\to 0$
it is the $F(4)$ theory that is relevant in this case.
As found in Section~\ref{sectfp} the eigenvalues at the flat f.p.\ with
$W=0$ are real only if
$d<11-2\sqrt{5}\approx 6.528$, thus $d$ must obey this bound.
Suppose there is some maximal value $d_m<6.528$ for which the $p=2$
solution runs into the RN f.p., then it must come arbitrarily close to
it for slightly larger values of $d$. As described above, the function $W$
should then stay small in a large $r$-interval, while $N$ runs from
$N\approx 0$ to $N\approx 1$. From Eqs.~(\ref{eigeng},\ref{eigenf})
it follows that in the considered domain for $d$ there are always two
convergent modes in the $(W,U)$ subsystem corresponding to two negative
eigenvalues for the RN f.p.\ ($\tau\to +\infty$) and two positive ones
for the flat f.p.\ ($\tau\to -\infty$).
However for solutions coming very close to
both f.p.s the modes with the eigenvalues of larger modulus are suppressed,
i.e.\ near the RN f.p.\ it is $\lambda_+^{\rm RN}$ that dominates, while it is
$\lambda_{-}^{\rm flat}$ at the flat f.p.
The best way to visualise the corresponding change of the $(W,U)$ system
moving from one f.p.\ to the other is to
consider the quotient $\eta=rU/W$ obeying a Riccati equation in the linear
approximation
\begin{equation}\label{Riccati}
\dot\eta=5-d+\Bigl((10-d)N-\kappa\Bigr)\eta-\eta^2\;,
\end{equation}
with $N$ and $\kappa$ given by the extremal RN solution
\begin{equation}
N=\Bigl(1-\frac{2M}{r^{d-3}}+\frac{d-5}{4(9-d)r^6}\Bigr)^{\frac{1}{2}}\;,
\quad {\rm with}\quad
M=\frac{3}{(9-d)}\Biggl(\frac{d-5}{4(d-3)}\Biggr)^{\frac{d-3}{6}}\;,
\end{equation}
and
\begin{equation}
\kappa=\frac{d-3+(5-d)N^2-\frac{d-5}{4r^6}}{2N}\;.
\end{equation}
Supposing there is a solution with the correct behaviour at both f.p.s we
should find a solution of the Riccati equation interpolating between
$\lambda_+^{\rm RN}$ at the RN f.p.\ ($r=r_0$) and $\lambda_-^{\rm flat}$
at the flat one ($r\to\infty$). For $d=6$ the Eq.~(\ref{Riccati}) can be
integrated
in closed form resulting in a solution running from $\lambda_+^{\rm RN}$
to $\lambda_+^{\rm flat}$
instead to $\lambda_-^{\rm flat}$,
excluding a bifurcation under the assumptions made.
For other values of $d$ in the relevant interval $5.632<d<6.528$
numerical integration of Eq.~(\ref{Riccati}) yields the same negative result.
Therefore we conclude that the dashed and solid curves in Fig.~\ref{sp1q2}
run back to different points at the $b$ axis, although numerically we
are unable to find regular solutions below $d\approx 6.8$.

There remains the interval $4<d<5$ for which no regular flat solutions
exist according to the virial theorem. On the other hand there are
self-gravitating solutions of the $F(2)$ model generalising the BM solutions
of $d=4$. In fact there are again families $b(d,\alpha)$ of regular
solutions tending to these BM type solutions in the limit $b\to 0$ and
$\alpha\to\infty$ after suitable rescalings
as discussed in Section~\ref{sectpq}.
Fig.~\ref{r1p1q2}
shows some of the curves $b(d,\alpha)$. For $b\to\infty$ they tend to
the conical f.p.\ as for $5\leq d<d_q$.

In order to further support our results for the 1-2-model we have also
performed some numerical calculations for the 2-3- and the 2-4-models.
Figs.~\ref{r1p2q3} and~\ref{r1p2q4} show again agreement with our analytical
results from Section~\ref{sectp}.
For $d<4p+1=9$ the curves $b(d,\alpha)$ have a BM type limit for
$\alpha\to\infty$ after suitable rescaling, while for $d>9$ they start
from $\alpha=0$ with the flat space solutions with finite values of $b$.
For $d<d_3\approx 8.63$ resp.\ $d<d_4\approx 11.81$ the curves approach a
finite limit for $\alpha$ as $b\to\infty$ and the solutions run
into the conical f.p., while for larger values of $d$ the curves
$b(d,\alpha)$ run back to $\alpha=0$ approaching the BM type solutions
of the $F(6)$ resp.\ $F(8)$ model after rescaling.
\mkfig{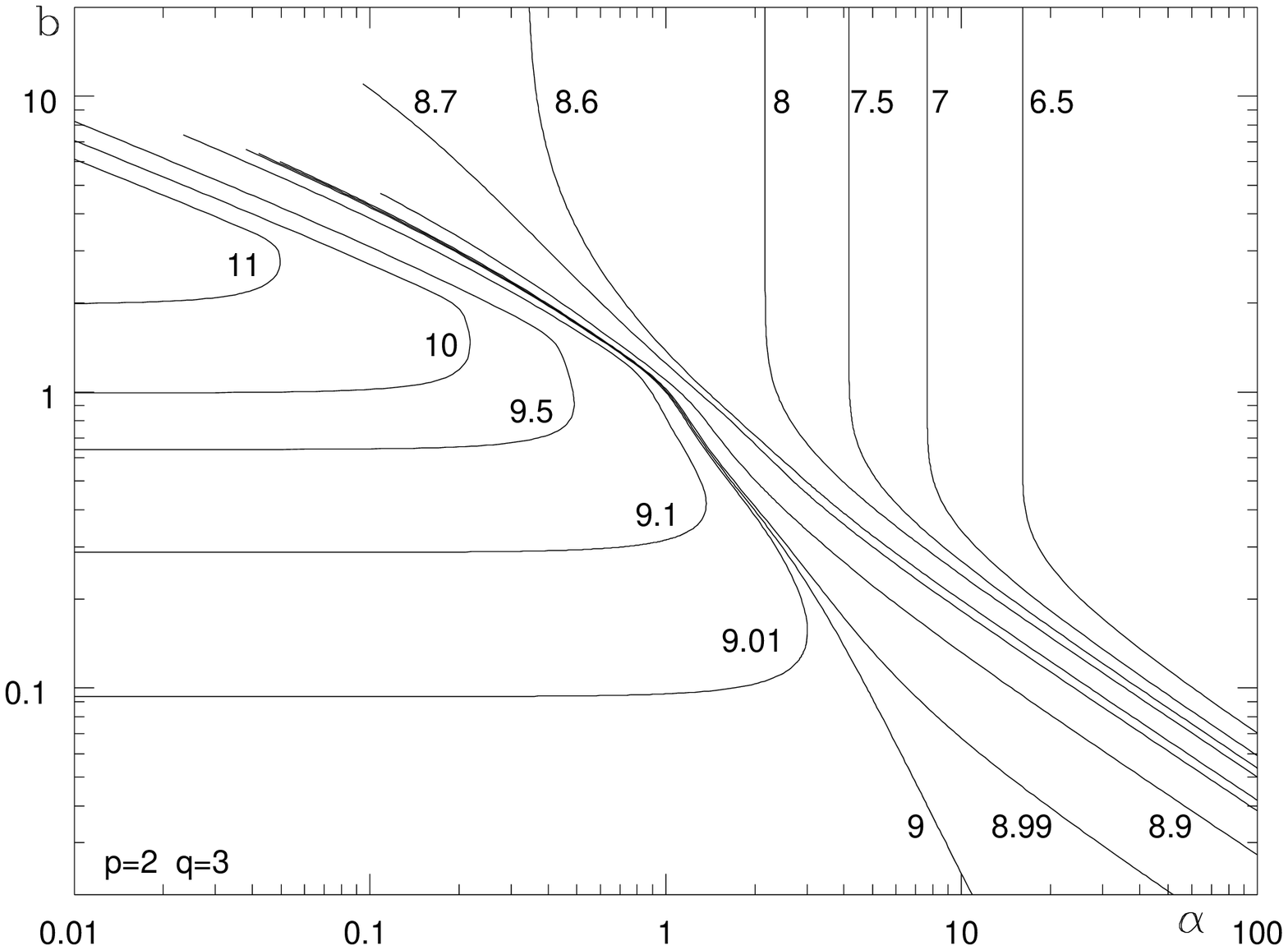}{0.9}{0.9}{45 178 561 566}
{$b$ vs.\ $\alpha$ for regular solutions of the 2-3-model with different
values of $d$}
\mkfig{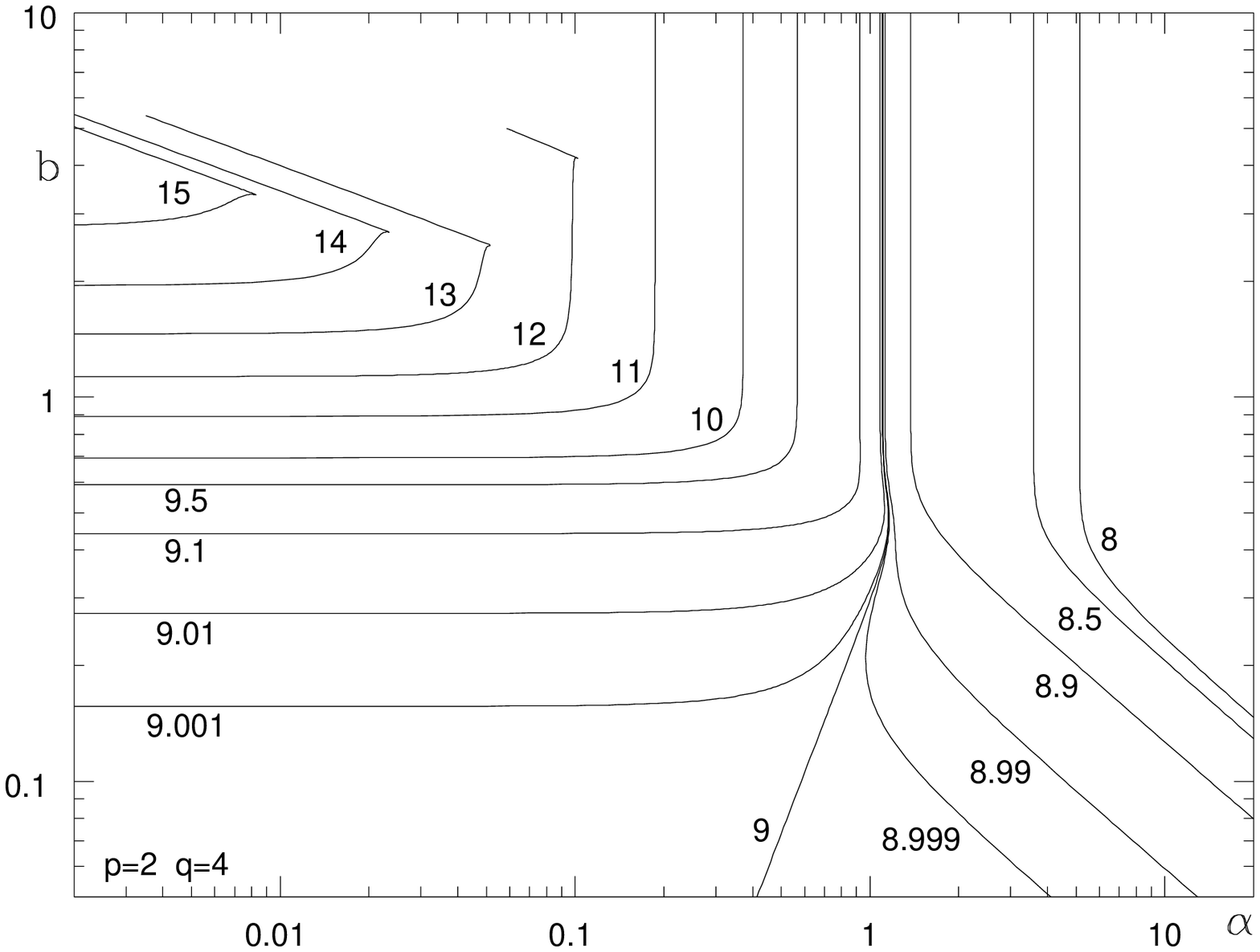}{0.9}{0.9}{45 178 561 566}
{$b$ vs.\ $\alpha$ for regular solutions of the 2-4-model with different
values of $d$}

The results on the f.p.s with $W=0$ and $N=0$ resp.\ $N=1$ have another
interesting consequence.
Solutions of the $p$-$q$-model, which come close to the RN f.p.\ and for which
$N\to 1$ for large $r$ come also close to the flat f.p.\ of the
$F(2p)$ model with $W=0$ (for $r\gg 1$ the $F(2p)$ terms dominate those of
the $F(2q)$ part).
Although the RN f.p.\ is non-oscillatory for $p>2$,
the flat one is oscillatory in the interval
$4p+3-2\sqrt{2p+1}<d<4p+3+2\sqrt{2p+1}$ with positive real part as long as
$d<4p+1$. Thus we expect to find regular solutions with any number of zeros
of $W$ accumulating at $r\to\infty$ for $4p+3-2\sqrt{2p+1}<d<4p+1$.
For $p=1$ this interval is $3.54<d<5$
(remember, however, that we assume $d\geq 4$), while
for $p=2$ we get $6.528<d<9$.
This is compatible with the observation of \cite{BCT} that no multi-node solutions
exist for the 1-2-model for $d\geq 5$. On the other hand for the 2-3-model
multi-node solutions are expected to exist for the integer dimensions $d=7$ and
$d=8$ and were actually found in our numerical analysis.
Figs.~\ref{r2p2d8} and~\ref{r3p2d8} show solutions of the $F(4)$ theory
for $d=8$ with two and three zeros of $W$.
Solutions for $d=7$ get much closer to the RN~f.p.\ and are therefore
more difficult to obtain numerically. Fig.~\ref{r2p2q3d7} shows a
solution of the 2-3-model for $d=7$ and $\alpha=20$ with two zeros of $W$.
\mkfig{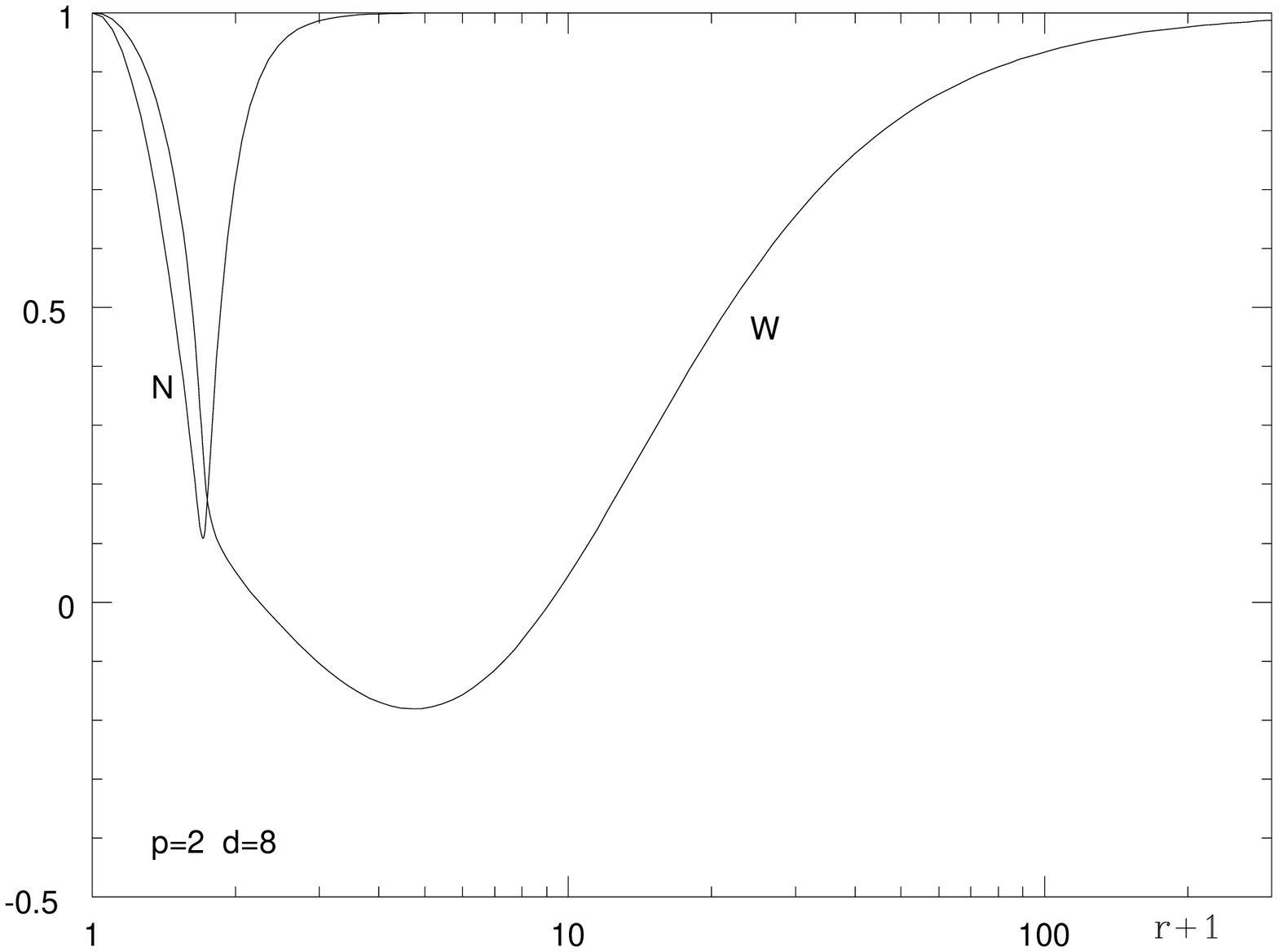}{0.9}{0.9}{45 178 561 566}
{$p=2$, $d=8$ solution with two zeros of $W$}
\mkfig{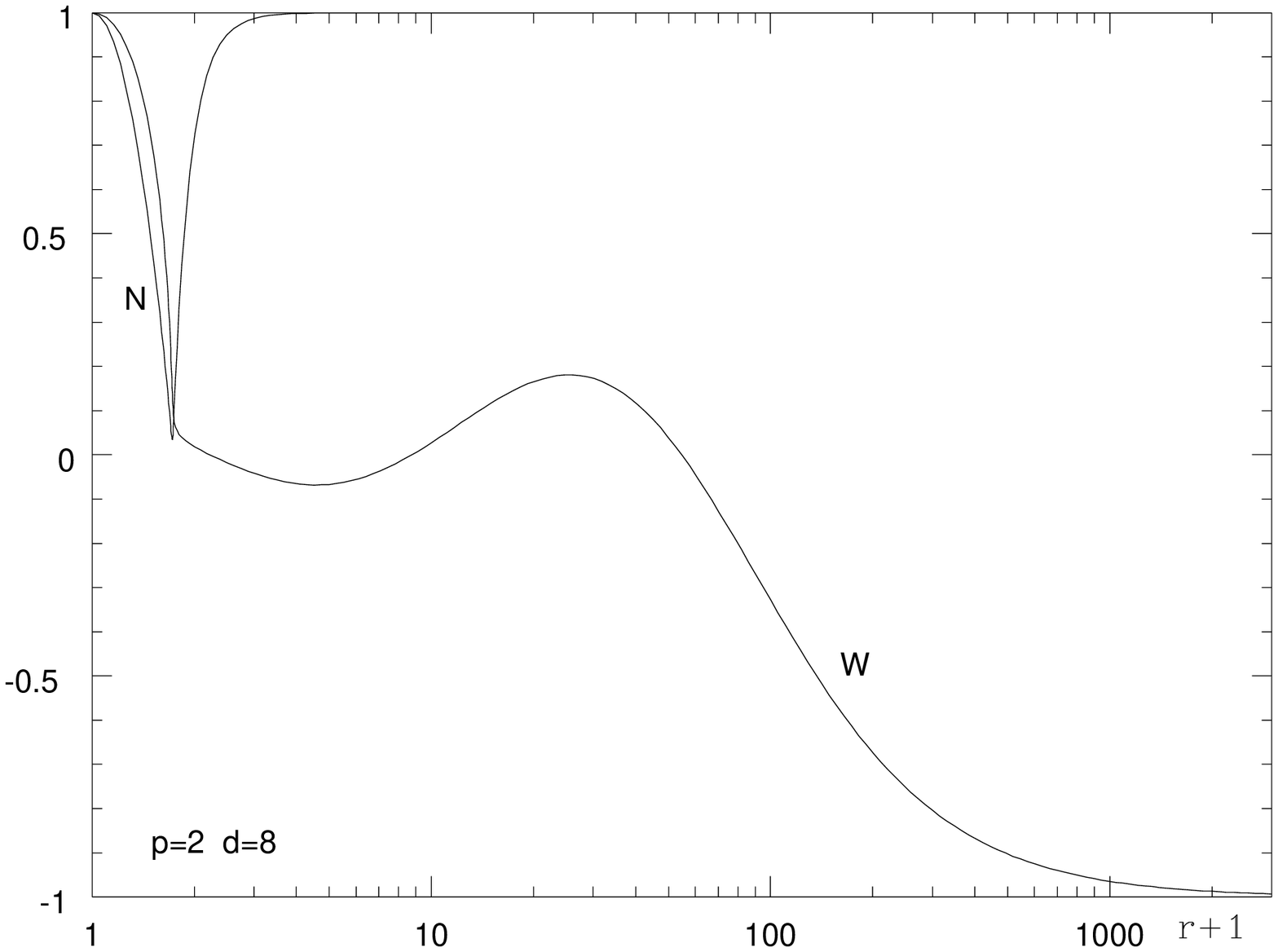}{0.9}{0.9}{45 178 561 566}
{$p=2$, $d=8$ solution with three zeros of $W$}
\mkfig{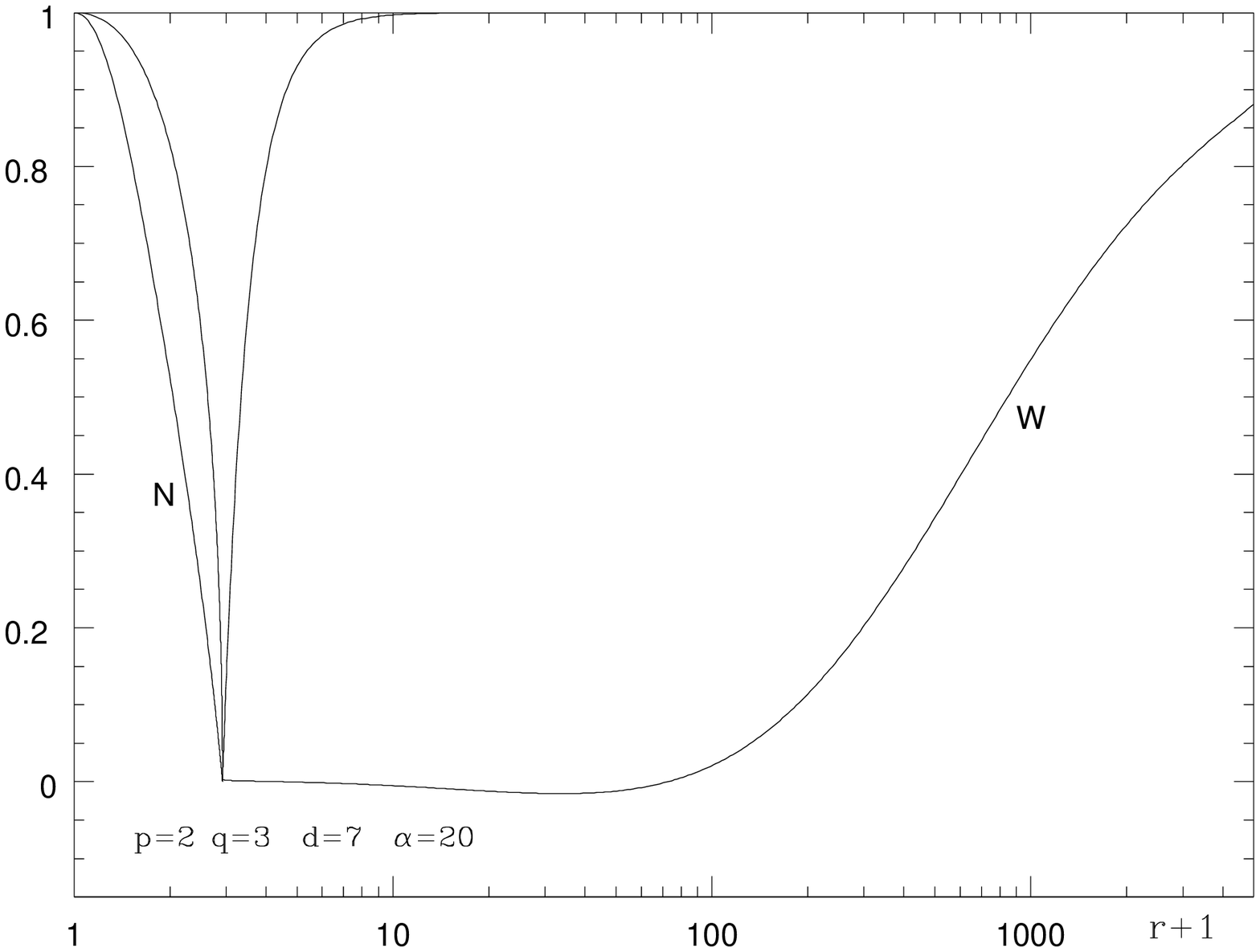}{0.9}{0.9}{45 178 561 566}
{Solution of the 2-3-model for $d=7$ and $\alpha=20$ with two zeros of $W$}

Since the eigenvalues of the f.p.\ have negative real part for $d>4p+1$
there are solutions with $W\to 0$ for $r\to\infty$
(see Fig.~\ref{p2inst}).
\mkfig{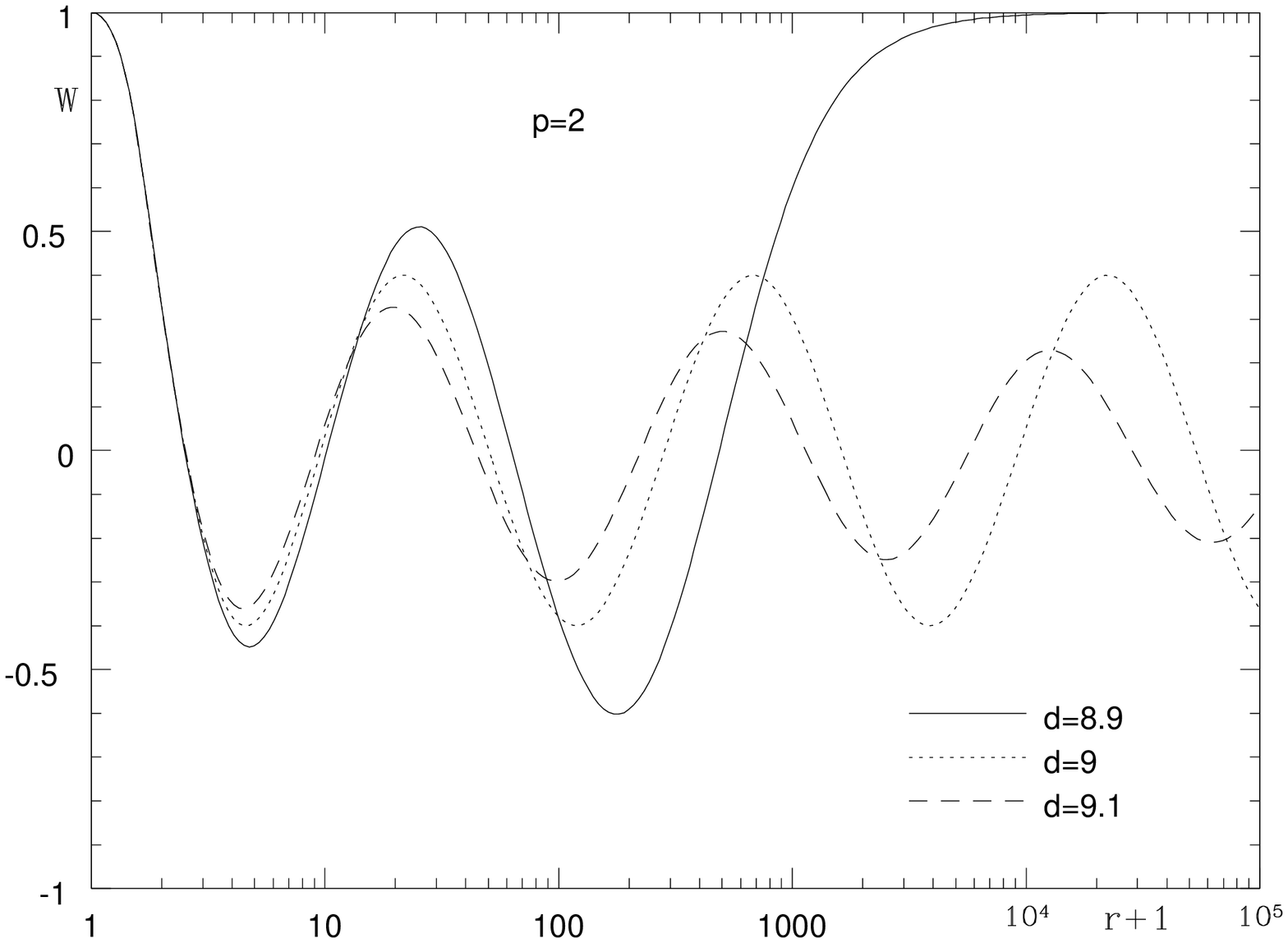}{0.9}{0.9}{45 178 561 566}
{Oscillating solutions of the $p=2$ model for $d=8.9$, $9$, and $9.1$ with
decreasing, constant, and growing amplitude respectively}

\section{Summary}
\label{conclusions}
Solutions to a family of EYM models in higher dimensions have been
studied. These models involve higher order YM curvature terms
$F(2p)$ characterised by integers $p$, $p=1$ giving the standard YM theory.
For reasons of simplicity we have restricted the number of such terms to two
($p$-$q$-models) depending on just one dimensionless parameter $\alpha$.
Studying the corresponding one-parameter families of (static, spherically symmetric)
solutions allows us to capture the qualitative properties of EYM
solutions in higher dimensions. We have restricted our study, for
simplicity, to regular solutions only, knowing that corresponding black
hole solutions exist.

The simplest example of the family of theories we have considered
(the 1-2-model) was recently
studied in \cite{BCT,BCHT} devised specifically to
suit spacetime dimensions $4<d<9$. The more general $p$-$q$-models considered here
suit also higher spacetime dimensions.
The central part of the work here was aimed at understanding qualitatively
the results found from the numerical analysis of \cite{BCT,BCHT}.
This was achieved by analytic analysis supplemented with numerical studies
including in addition to the previously considered case $p=1$, $q=2$ also
the cases $p=2$, $q=3$ and $p=2$, $q=4$.
The analytic part of the work was that of a fixed point
analysis of nonlinear ordinary differential equations,
employed previously \cite{BFM,BFM1} for the familiar $d=4$ BM solutions
and magnetic monopoles.

Like in the case of the gravitating monopoles \cite{BFM1} of the EYMH system,
one finds families of solutions parametrised by the
dimensionless parameter $\alpha$, which can be understood as a quotient of
mass scales of the YM and the gravity parts of the theory.
Since the $\alpha$ dependence of the solutions shows several distinctive
behaviours for different dimensions it turned out to be particularly
useful to vary the dimension parameter $d$ continously, which is possible
due to the explicit dependence of the radial field equations on $d$.
This trick allowed us to relate the changes with $d$ to changes in
the fixed point structure of the equations.
We find several different possibilities for the $\alpha$ dependence of the
solutions that can be characterised by their behaviour for small $b$ and for
large $b$ (see Figs.~\ref{r1p1q2}, \ref{r1p2q3}, and~\ref{r1p2q4}).
There are three possibilities for small $b$:
\begin{itemize}
\item
For $2q<d<4p+1$ the parameter $\alpha$ decreases from infinity as $b$
increases from zero. In the limit $\alpha\to\infty$ the solutions approach a
suitably rescaled BM type solution of the $F(2p)$ model.
\item
For $d=4p+1$, as $b$ starts from zero the parameter $\alpha$ either
increases from zero if $4p<2q+1$, or decreases from infinity if $4p>2q+1$.
In the limit $b\to0$ the solutions approach the instanton of the flat
$F(2p)$ model.
\item
For $4p+1<d<4q+1$ the parameter $\alpha$ increases from zero as $b$
starts from a finite value. For $\alpha\to 0$ the solutions approach a
solution of the flat $p$-$q$-model.
\end{itemize}
And there are two possibilities for large $b$:
\begin{itemize}
\item
For $2q<d\le d_q$ the parameter $\alpha$ approaches a finite limiting value
$\alpha_c\neq 0$ as $b\to\infty$. In the limit the solutions tend to a new
type of fixed point, which we named `conical' in view of the
corresponding space-time singularity.
\item
For $d_q<d<4q+1$ the parameter $\alpha$ decreases to zero as $b\to\infty$.
For $\alpha\to0$ the solutions approach the BM type solution of the $F(2q)$
model. Note, however, that this limit cannot be explored numerically for
values of $d$ close to $d_q$ because the solutions come too close to the RN
f.p.
\end{itemize}
Clearly not all six combinations of these possibilities can be realised
simultaneously for one $p$-$q$-model, because some of the constraints on
them are mutually exclusive.

The existence of multi-node solutions is related to the
existence of a f.p.\ with complex eigenvalues corresponding to a focal
point. In the present case there is such a f.p.\ with vanishing
YM potential $W$.
Our analytical analysis reveals that for the $p$-$q$-models
multi-node solutions are expected for dimensions in the interval
$4p+3-2\sqrt{2p+1}<d<4p+1$ perfectly compatible with our numerical results.

\section*{Acknowledgements}
Thanks to Eugen Radu for useful discussions. We are grateful
to the Alexander-von-Humboldt Foundation and the Dublin Institute
for Advanced Studies for their support. This work was carried out
in the framework of projects SC/03/390 and IC/05/03 of
Enterprise-Ireland.

\newpage

\mkfigs


\newpage

\begin{thebibliography}{99}

\newcommand{\AP}{\sl Annals  Phys.\ \bf}
\newcommand{\CMP}{\sl Commun.\ Math.\ Phys.\ \bf}
\newcommand{\CQG}{\sl Class.\ Quantum\ Grav.\ \bf}
\newcommand{\FP}{\sl Fortsch.\ Phys.\ \bf}
\newcommand{\JETPL}{\sl JETP Lett.\ \bf}
\newcommand{\JHEP}{\sl JHEP\ \bf}
\newcommand{\JMP}{\sl J. Math.\ Phys.\ \bf}
\newcommand{\JPA}{\sl J. Phys.\ \bf A\,}
\newcommand{\NPB}{\sl Nucl.\ Phys.\ \bf B\,}
\newcommand{\NPPS}{\sl Nucl.\ Phys.\ Proc.\ Suppl.\ \bf}
\newcommand{\PLB}{\sl Phys.\ Lett.\ \bf B\,}
\newcommand{\PRD}{\sl Phys.\ Rev.\ \bf D\,}
\newcommand{\PRL}{\sl Phys.\ Rev.\ Lett.\ \bf}

\bibitem{BM}
R. Bartnik and J. McKinnon, {\PRL 61} (1988) 141.

\bibitem{VG}
M.S. Volkov and D.V. Gal'tsov, {\JETPL 50} (1989) 346.

\bibitem{B}
P. Bizon, {\PRL 64} (1990) 2844.

\bibitem{Kuenzle}
H.P.~K\"unzle and A.K.M.~Masood ul Alam,
{\JMP 31} (1990) 928.

\bibitem{GSW}
M.B. Green, J.H. Schwarz and E. Witten, {\it Superstring Theory},
Cambridge University Press, Cambridge, 1987.

\bibitem{HS}
J. Harvey and A. Strominger, {\it TASI lectures on quantum aspects of black
holes}, hep-th/9209055.

\bibitem{Pol}
J. Polchinski, {\it TASI lectures on D-branes}, hep-th/9611050.

\bibitem{BCT}
Y. Brihaye, A. Chakrabarti and D.H. Tchrakian,
{\CQG 20} (2003) 2765 [hep-th/0202141].

\bibitem{BCHT}
Yves Brihaye, A. Chakrabarti, Betti Hartmann and D.H. Tchrakian,
{\PLB 561} (2003) 161 [hep-th/0212288].

\bibitem{Tseytlin}
A.A. Tseytlin, {\it Born--Infeld action, suersymmetry and string theory},
in Yuri Golfand memorial volume, ed. M. Shifman, World Scientific, 2000.

\bibitem{BRS}
E. Bergshoeff, M. de Roo and A. Sevrin,
{\FP 49} (2001) 433-440; {\NPPS 102} (2001) 50-55.

\bibitem{CNT}
M. Cederwall, B. Nilsson and D. Tsimpis, {\JHEP 0106} (2001) 034.

\bibitem{BFM1}
P. Breitenlohner, P. Forgacs and D. Maison,
{\NPB 383} (1992) 357; {\it ibid.} {\bf 442} (1995) 126.

\bibitem{BFM}
P.~Breitenlohner, P.~Forg\'acs and D.~Maison,
{\CMP 163} (1994) 141

\bibitem{Hart}
P.~Hartman, {\it Ordinary Differential Equations},
Boston: Birkh\"auser, 1982.

\bibitem{barriola}
M.~Barriola and A.~Vilenkin,
{\PRL 63} (1989) 341.

\bibitem{butc}
J. Burzlaff and D. H. Tchrakian, {\JPA 26} (1993) L1053.

\bibitem{BFMmon}
P. Breitenlohner, P. Forgacs and D. Maison,
{\NPB 442} (1995) 126.

\end{thebibliography}
\end{document}